\documentclass[aps,prd,preprint,nofootinbib,showpacs,showkeys,preprintnumbers,amsmath,amssymb,11pt]{revtex4}

\usepackage{graphicx}
\usepackage{dcolumn}
\usepackage{bm}
\usepackage{bbm}

\usepackage{lscape}

\allowdisplaybreaks

\begin{document}


\title{Neutrino Oscillation Probabilities in Matter with \emph{Direct} and \emph{Indirect} Unitarity Violation in the Lepton Mixing Matrix\vspace{0.4cm}}

\author{\bf Yu-Feng Li}
\email{liyufeng@ihep.ac.cn}
\affiliation{Institute of High Energy Physics, Chinese Academy of Sciences, P.O. Box 918, Beijing 100049, China}

\author{\bf Shu Luo}
\email{luoshu@xmu.edu.cn}
\affiliation{Department of Astronomy and Institute of Theoretical Physics and Astrophysics, Xiamen University, Xiamen, Fujian 361005, China\vspace{1cm}}

\begin{abstract}
\vspace{0.3cm}
\noindent
In the presence of both \emph{direct} and \emph{indirect} unitarity violation in the lepton mixing matrix, we derive a complete set of series expansion formulas for neutrino oscillation probabilities in matter of constant density. Expansions in the mass hierarchy parameter $\alpha \equiv \Delta m_{21}^{2} / \Delta m_{31}^{2}$ and those unitarity violation parameters $s^{2}_{ij}$ (for $i = 1, 2, 3$ and $j = 4, 5, 6$) up to the first order are studied in this paper. We analyse the accuracy of the analytical series expansion formulas in different regions of $L / E$. A detailed numerical analysis is also performed, of which the different effects of the direct and the indirect unitarity violation are particularly emphasized. We also study in this paper the summed $\nu^{}_{\alpha} \rightarrow \nu^{}_{e, \nu, \tau}$ probabilities, whose deviation from the unity provides a definite signal of the unitarity violation.
\vspace{0.3cm}
\end{abstract}



\maketitle

\section{Introduction}

The standard three-flavor paradigm of neutrino mixing has been solidly established by the solar, atmospheric, reactor and accelerator neutrino experiments \cite{PDG}. However, the anomalies from the LSND \cite{LSND}, MiniBooNE \cite{MiniBooNE}, reactor antineutrino \cite{reactor} and Gallium radioactive source \cite{Gallium} experiments as well as the analysis of current cosmological observations \cite{Cosmology} have provided interesting hints that the three-flavor framework may be incomplete and there may exist additional light sterile neutrinos \cite{sterile,stefit}. From the theoretical point of view, sterile neutrinos are regarded as natural ingredients of different types of seesaw models \cite{seesaw} as well as warm dark matter candidates \cite{keV}. Nevertheless, the number of sterile neutrino species, the neutrino mass scale as well as the patten of active-sterile mixing are all currently unknown, which consistent one of the most important tasks for future precision neutrino experiments.

The mixing of three active neutrinos $\nu^{}_e$, $\nu^{}_\mu$ and $\nu^{}_\tau$ with the possible existing sterile ones could result in unitarity violation (UV) \cite{UVtest,Antusch2006,Antusch2014,Escrihuela:2015wra,Parke:2015goa} of the $3 \times 3$ Maki-Nakagawa-Sakata-Pontecorvo (MNSP) lepton mixing matrix \cite{MNSP}, where the sterile neutrinos can be either heavy or light. We call the former case \emph{indirect} unitarity violation (IUV) and the latter \emph{direct} unitarity violation (DUV) \cite{UVtest}. These two kinds of unitarity violation may have distinct effects in neutrino oscillation experiments, where the reason is that heavy\footnote{Our definition of ``heavy'' sterile neutrinos assumes that they cannot be produced via the weak interaction process as a result of energy conservation. Therefore, the minimal masses of the heavy neutrinos are process dependent. For illustration,  in the on-shell $W^{+}_{}$ decay, $W^{+}_{} \rightarrow l^{+}_{} + \nu^{}_{l}$, the maximal mass of neutrinos which can be produced is $\sqrt{m^{2}_{W}-m^{2}_{l}}$. In this particular process the definition of "heavy" is $> \sqrt{m^{2}_{W}-m^{2}_{l}}$.}
massive sterile neutrinos are kinematically forbidden in the neutrino production and detection processes of low energy neutrino experiments, while light sterile neutrinos are able to participate in neutrino oscillations as their active partners. The purpose of this paper is to derive a complete set of approximate analytical formulas for neutrino oscillation probabilities in matter in the presence of both the light and the heavy sterile neutrinos. The different effects of the DUV and the IUV on neutrino oscillation experiments are particularly emphasized. A complete numerical calculation of these probabilities is also presented to make a comparative study.

This paper is organized as follows. In Sec.~II, we briefly introduce the general framework of the ($3$+$\mathbbm{1}$+$\mathbf{2}$) scenario, which is essential for studying the UV effects. The different effects of the DUV and the IUV are discussed. In Sec.~III, we present the series expansion formulas for the neutrino oscillation probabilities in matter of constant density up to the first order in both the mass hierarchy parameter $\alpha \equiv \Delta m_{21}^{2} / \Delta m_{31}^{2}$ and those UV parameters $s^{2}_{ij}$ (for $i = 1, 2, 3$, and $j = 4, 5, 6$). The corresponding probabilities in vacuum up to the same order are also presented. Section IV is attributed to the numerical calculations. We show in this section the qualitative behaviors of neutrino oscillation probabilities, and test the accuracy of these analytical formulas. Finally, we summarise in Sec.~V, and the details of the perturbative expansion of the neutrino oscillation probabilities in the case of UV are given in the appendices.

\section{General Framework in the Presence of Unitarity Violation}

In this paper, we consider the ($3$+$\mathbbm{1}$+$\mathbf{2}$) scenario with three additional species of sterile neutrinos, of which one is the light sterile neutrino and the other two are heavy sterile ones
\footnote{Current cosmological observations favor at most one species of light sterile neutrinos \cite{Cosmology}.},
and discuss the possibility of determining both the DUV and the IUV parameters in neutrino oscillation experiments.

\subsection{Parametrization of the Lepton Mixing Matrix}

In the ($3$+$\mathbbm{1}$+$\mathbf{2}$) scenario, the full picture of neutrino mixing are described by a $6 \times 6$ unitary matrix ${\cal U}$,
which can be decomposed as \cite{Xing2012,Dev:2012bd}
\begin{eqnarray}
{\cal U} \; = \; \left ( \begin{matrix} {\bm 1} & {\bm 0} \cr {\bm 0} & U^{}_{0} \cr \end{matrix} \right ) \left ( \begin{matrix} A & R \cr S & B \cr \end{matrix} \right ) \left ( \begin{matrix} V^{}_{0} & {\bm 0} \cr {\bm 0} & {\bm 1} \cr \end{matrix} \right ) \; = \; \left ( \begin{matrix} A V^{}_{0} & R \cr U^{}_{0} S V^{}_{0} & U^{}_{0} B \cr \end{matrix} \right ) \; ,
\end{eqnarray}
where $V^{}_{0}$ and $U^{}_{0}$ are unitary matrices while $A$, $B$, $R$ and $S$ are not unitary, $\bm{0}$ and $\bm{1}$ stand for the zero and identity matrices respectively. All of them are $3 \times 3$ matricies. The explicit expressions of these matrices can be found in Ref.~\cite{Xing2012}.  The full $6 \times 6$ unitary mixing matrix ${\cal U}$ is parameterized by altogether 15 mixing angles and 15 CP-violating phases (including the Majorana phases).

In the presence of three additional sterile neutrinos, the Lagrangian of the standard weak charged-current interaction should be written as
\begin{eqnarray}
- {\cal L}^{}_{CC} = \frac{g}{\sqrt{2}} ~ \overline{ \left( \begin{matrix} e & \mu & \tau \end{matrix} \right)^{}_{\rm L} } ~ \gamma^{\mu}_{} \left[ V \left( \begin{matrix} \nu^{}_1 \cr \nu^{}_2 \cr \nu^{}_3 \end{matrix} \right)^{}_{\rm L} + R \left( \begin{matrix} \nu^{}_4 \cr \nu^{}_5 \cr \nu^{}_6 \end{matrix} \right)^{}_{\rm L} \right] W^{-}_{\mu} + {\rm h.c.} \; ,
\end{eqnarray}
where $\nu^{}_{4}$ is the mass eigenstate of the light sterile neutrino and $\nu^{}_{5}$, $\nu^{}_{6}$ are mass eigenstates of the two heavy sterile neutrinos.
$V \equiv A V^{}_{0}$ is just the MNSP matrix, which is in general non-unitary in this scenario. Equation (2) indicates that the sterile neutrinos can participate in the charged-current interaction through their mixing with the active ones (which is described by the mixing matrix $R$) and therefore result in low energy signals of the UV.
The matrix $V^{}_{0}$ in Eq.~(1) can be parametrised using the standard parametrization as \cite{PDG}
\begin{eqnarray}
V^{}_{0} \; = \; \left ( \begin{matrix} c^{}_{12} c^{}_{13} & s^{}_{12} c^{}_{13} & s^{}_{13} e^{-i\delta}_{} \cr -s^{}_{12} c^{}_{23} - c^{}_{12} s^{}_{23} s^{}_{13} e^{i\delta}_{} & c^{}_{12} c^{}_{23} - s^{}_{12} s^{}_{23} s^{}_{13} e^{i\delta}_{} & s^{}_{23} c^{}_{13} \cr s^{}_{12} s^{}_{23} - c^{}_{12} c^{}_{23} s^{}_{13} e^{i\delta}_{} & - c^{}_{12} s^{}_{23} - s^{}_{12} c^{}_{23} s^{}_{13} e^{i\delta}_{} & c^{}_{23} c^{}_{13} \end{matrix} \right ) \; .
\end{eqnarray}
To the order of $s^{2}_{ij}$ ($i = 1, 2, 3\; {\rm and}\; j = 4, 5, 6$), $A$ and $R$ can be approximately written as \cite{Xing2012}
\begin{eqnarray}
A & \simeq & {\bm 1} - \left( \begin{matrix} \displaystyle\frac{1}{2} \left( s^2_{14} + s^2_{15} + s^2_{16} \right) & 0 & 0 \cr \hat{s}^{}_{14} \hat{s}^*_{24} + \hat{s}^{}_{15} \hat{s}^*_{25} + \hat{s}^{}_{16} \hat{s}^*_{26} & \displaystyle\frac{1}{2} \left( s^2_{24} + s^2_{25} + s^2_{26} \right) & 0 \cr \hat{s}^{}_{14} \hat{s}^*_{34} + \hat{s}^{}_{15} \hat{s}^*_{35} + \hat{s}^{}_{16} \hat{s}^*_{36} & \hat{s}^{}_{24} \hat{s}^*_{34} + \hat{s}^{}_{25} \hat{s}^*_{35} + \hat{s}^{}_{26} \hat{s}^*_{36} & \displaystyle\frac{1}{2} \left( s^2_{34} + s^2_{35} + s^2_{36} \right) \end{matrix} \right) \; , \\
R & \simeq & \left( \begin{matrix} \hat{s}^*_{14} & \hat{s}^*_{15} & \hat{s}^*_{16} \cr \hat{s}^*_{24} & \hat{s}^*_{25} & \hat{s}^*_{26} \cr \hat{s}^*_{34} & \hat{s}^*_{35} & \hat{s}^*_{36} \end{matrix} \right) \; .
\end{eqnarray}
Here $s^{}_{ij} \equiv \sin\theta^{}_{ij}$, $c^{}_{ij} \equiv \cos\theta^{}_{ij}$ and $\hat{s}^{}_{ij} \equiv e^{i \delta^{}_{ij}}_{} \sin\theta^{}_{ij}$ with $\theta^{}_{ij}$ and $\delta^{}_{ij}$ being the rotation and phase angles, respectively. We can find that there are altogether 9 additional mixing angles and 9 additional phases in $R$ and $A$, which are relevant to the low energy experiments. These UV parameters can be divided into two groups:
\begin{itemize}
\item  ($\theta^{}_{14}$, $\theta^{}_{24}$, $\theta^{}_{34}$) and ($\delta^{}_{14}$, $\delta^{}_{24}$ $\delta^{}_{34}$) are parameters of the DUV describing the mixing between three active neutrinos and the light sterile neutrino. Since the light sterile neutrino can be produced and detected in the low energy experiments, these parameters will also appear in the neutrino oscillation probabilities as those standard mixing parameters.
\item  ($\theta^{}_{15}$, $\theta^{}_{16}$, $\theta^{}_{25}$, $\theta^{}_{26}$, $\theta^{}_{35}$, $\theta^{}_{36}$) and ($\delta^{}_{15}$, $\delta^{}_{16}$, $\delta^{}_{25}$, $\delta^{}_{26}$, $\delta^{}_{35}$, $\delta^{}_{36}$) are the IUV parameters that describe the mixing between three active neutrinos and the heavy sterile ones. The IUV affects the processes of neutrino production and detection, and therefore induce the ``zero-distance'' effects \cite{zero} in neutrino oscillations.
\end{itemize}
The analyses in Refs.~\cite{Antusch2006,Antusch2014} show that leptonic and semileptonic decays are quite sensitive to the IUV of the leptonic mixing matrix. The strongest constraints come from the rare charged lepton decay experiments, which indicate that the combinations $\left ( U U^{\dagger}_{} \right )^{}_{\alpha \beta} \lesssim 1 \%$, with $\alpha, \beta = e, \mu, \tau$, should be satisfied
\footnote{Note that, instead of the $3 \times 3$ non-unitary MNSP matrix $V$, the matrix $U$ is the $3 \times 4$ left-up sub-matrix of $\cal U$ as showed in Eq.~(6). Since the light sterile neutrino can be kinematically produced in the electroweak decays, the DUV parameters are irrelevant to the strong constraints obtained there~\cite{Antusch2006,Antusch2014}.}.
On the other hand, the DUV parameters can be revealed in the very-short-baseline neutrino oscillation experiments. The global analysis of current results of the short-baseline neutrino experiments points towards the eV scale light sterile neutrino(s) with active-sterile mixing parameters $|U^{}_{\alpha 4}|^2$ ($\alpha = e, \mu, \tau$) of a few percent \cite{stefit}. In this paper, we simply take all the active-sterile mixing angles $\theta^{}_{ij}$ ($i = 1, 2, 3$ and $j = 4, 5, 6$) as small parameters of the same order $s$.

\subsection{Neutrino Oscillation in the Presence of Unitarity Violation}

In this paper, we restrict us to the typical neutrino oscillation process $\nu^{}_{\alpha} \rightarrow \nu^{}_{\beta}$ where both the production of $\nu^{}_{\alpha}$ and the detection of $\nu^{}_{\beta}$ are via the charged-current interaction. Since in our proposed ($3$+$\mathbbm{1}$+$\mathbf{2}$) scenario, there is only one light sterile neutrino that can participate in the neutrino oscillation together with the three active ones, only the elements in the $3 \times 4$ left-up sub-matrix $U$ of the full $6 \times 6$ mixing matrix $\cal U$ are related to the neutrino oscillation probabilities:
\begin{eqnarray}
\left ( \begin{matrix} \nu^{}_{e} \cr \nu^{}_{\mu} \cr \nu^{}_{\tau} \end{matrix} \right ) \; = \; U \left ( \begin{matrix} \nu^{}_{1} \cr \nu^{}_{2} \cr \nu^{}_{3} \cr \nu^{}_{4} \end{matrix} \right ) \; = \; \left ( \begin{matrix} U^{}_{e1} & U^{}_{e2} & U^{}_{e3} & U^{}_{e4} \cr U^{}_{\mu 1} & U^{}_{\mu 2} & U^{}_{\mu 3} & U^{}_{\mu 4} \cr U^{}_{\tau 1} & U^{}_{\tau 2} & U^{}_{\tau 3} & U^{}_{\tau 4} \end{matrix} \right ) \; \left ( \begin{matrix} \nu^{}_{1} \cr \nu^{}_{2} \cr \nu^{}_{3} \cr \nu^{}_{4} \end{matrix} \right ) \; .
\end{eqnarray}
In this case, the neutrino oscillation probability in vacuum should be written as \cite{Antusch2006,Luo2008}
\begin{eqnarray}
P \; ( \stackrel{(-)}{\nu}^{}_{\alpha} \rightarrow  \stackrel{(-)}{\nu}^{}_{\beta} ) & = & \frac{\left | \left ( U^{*}_{} e_{}^{-i E L} U^{T}_{} \right )_{\alpha \beta} \right |^2}{\left ( U_{}^{} U_{}^{\dagger} \right )_{\alpha \alpha} \left ( U_{}^{} U_{}^{\dagger} \right )_{\beta \beta}} \; = \; \frac{\left | \displaystyle \sum^{}_{i=1,2,3,4} \left ( U^{*}_{\alpha i} \; e_{}^{-i \frac{m^{2}_{i} L}{2E^{}_{\nu}}} \; U^{}_{\beta i} \right )_{\alpha \beta} \right |^2}{\Big ( \displaystyle \sum^{}_{i=1,2,3,4} | U^{}_{\alpha i} |^2 \Big ) \Big ( \displaystyle \sum^{}_{i=1,2,3,4} | U^{}_{\beta i} |^2 \Big )} \nonumber\\
& = & \frac{1}{ \Big ( \displaystyle \sum^{}_{i=1,2,3,4} | U^{}_{\alpha i} |^2 \Big ) \Big ( \displaystyle \sum^{}_{i=1,2,3,4} | U^{}_{\beta i} |^2 \Big )} \Bigg \{ \left | \sum^{}_{i=1,2,3,4} U^{*}_{\alpha i} U^{}_{\beta i} \right |^2 \nonumber\\
& & - 4 \sum^{}_{j > i} {\rm Re} \left [ U^{}_{\alpha i} U^{}_{\beta j} U^{*}_{\alpha j} U^{*}_{\beta i} \right ] \sin^2 \Delta^{}_{ji} \pm 2 \sum^{}_{j > i} {\rm Im} \left [ U^{}_{\alpha i} U^{}_{\beta j} U^{*}_{\alpha j} U^{*}_{\beta i} \right ] \sin2\Delta^{}_{ji} \Bigg \} \nonumber\\
& = & \frac{1}{ \Big ( \displaystyle \sum^{}_{i=1,2,3,4} | U^{}_{\alpha i} |^2 \Big ) \Big ( \displaystyle \sum^{}_{i=1,2,3,4} | U^{}_{\beta i} |^2 \Big )} \Bigg \{ \left | \sum^{}_{i=1,2,3,4} U^{*}_{\alpha i} U^{}_{\beta i} \right |^2 \nonumber\\[1mm]
& & - 4 {\rm Re} \left [ U^{}_{\alpha 1} U^{}_{\beta 2} U^{*}_{\alpha 2} U^{*}_{\beta 1} \right ] \sin^2 \Delta^{}_{21} \pm 2 {\rm Im} \left [ U^{}_{\alpha 1} U^{}_{\beta 2} U^{*}_{\alpha 2} U^{*}_{\beta 1} \right ] \sin2\Delta^{}_{21} \nonumber\\[1mm]
& & - 4 {\rm Re} \left [ U^{}_{\alpha 1} U^{}_{\beta 3} U^{*}_{\alpha 3} U^{*}_{\beta 1} \right ] \sin^2 \Delta^{}_{31} \pm 2 {\rm Im} \left [ U^{}_{\alpha 1} U^{}_{\beta 3} U^{*}_{\alpha 3} U^{*}_{\beta 1} \right ] \sin2\Delta^{}_{31} \nonumber\\[1mm]
& & - 4 {\rm Re} \left [ U^{}_{\alpha 2} U^{}_{\beta 3} U^{*}_{\alpha 3} U^{*}_{\beta 2} \right ] \sin^2 \Delta^{}_{32} \pm 2 {\rm Im} \left [ U^{}_{\alpha 2} U^{}_{\beta 3} U^{*}_{\alpha 3} U^{*}_{\beta 2} \right ] \sin2\Delta^{}_{32} \nonumber\\[1mm]
& & - 4 {\rm Re} \left [ U^{}_{\alpha 1} U^{}_{\beta 4} U^{*}_{\alpha 4} U^{*}_{\beta 1} \right ] \sin^2 \Delta^{}_{41} \pm 2 {\rm Im} \left [ U^{}_{\alpha 1} U^{}_{\beta 4} U^{*}_{\alpha 4} U^{*}_{\beta 1} \right ] \sin2\Delta^{}_{41} \nonumber\\[1mm]
& & - 4 {\rm Re} \left [ U^{}_{\alpha 2} U^{}_{\beta 4} U^{*}_{\alpha 4} U^{*}_{\beta 2} \right ] \sin^2 \Delta^{}_{42} \pm 2 {\rm Im} \left [ U^{}_{\alpha 2} U^{}_{\beta 4} U^{*}_{\alpha 4} U^{*}_{\beta 2} \right ] \sin2\Delta^{}_{42} \nonumber\\[0mm]
& & - 4 {\rm Re} \left [ U^{}_{\alpha 3} U^{}_{\beta 4} U^{*}_{\alpha 4} U^{*}_{\beta 3} \right ] \sin^2 \Delta^{}_{43} \pm 2 {\rm Im} \left [ U^{}_{\alpha 3} U^{}_{\beta 4} U^{*}_{\alpha 4} U^{*}_{\beta 3} \right ] \sin2\Delta^{}_{43} \; \Bigg \} \; ,
\end{eqnarray}
where $\Delta^{}_{ji} \equiv \Delta m^2_{ji} L / 4E$ with $\Delta m^2_{ji} \equiv m^{2}_{j} - m^{2}_{i}$ being the neutrino mass-squared difference. Here the Greek letters $\alpha$, $\beta$ are the flavor indices $e$, $\mu$, $\tau$, while the Latin letters $i$, $j$ are the indices of mass eigenstates. Note that the indices $i$, $j$ run over only the light neutrinos (both the active and sterile ones) which can be kinematically produced in neutrino oscillation experiments, and the normalization factor $1 /( \sum^{}_{i=1,2,3,4} | V^{}_{\alpha i} |^2 ) ( \sum^{}_{i=1,2,3,4} | V^{}_{\beta i} |^2 )$ ensures that at the source we have $\sum_{i}^{} |A(W^{+}_{} \rightarrow \bar{l}_{\alpha}^{} \nu_{i}^{})|^2 = 1$ and at the detector $\sum_{i}^{} |A(\nu_{i}^{} W^{-}_{} \rightarrow l_{\beta}^{})|^2 = 1$ is satisfied.

When neutrinos propagate through the Earth before reaching the detector, neutrinos may interact with electrons, protons and neutrons in the medium via the charged-current (CC) interactions or neutral-current (NC) interactions. The coherent forward scattering from the constituents of the Earth matter will modify the evolution behavior of neutrinos. For the propagation of neutrinos in matter, the Hamiltonian can be written in the basis of mass eigenstates in vacuum as
\begin{eqnarray}
\tilde{\cal{H}} \; = \; E + U^{T}_{} \bar{A} U^{*}_{} \; ,
\end{eqnarray}
where $E \equiv {\rm diag} \left \{ E^{}_{1}, E^{}_{2}, E^{}_{3}, E^{}_{4} \right \}$ is the energy matrix, $\bar{A} \equiv {\rm diag} \left \{ V^{}_{CC} - V^{}_{NC}, \; - V^{}_{NC}, \; - V^{}_{NC} \right \}$ with $V_{CC}^{} \equiv \sqrt{2} G_{F}^{} n_{e}^{}$ and $V_{NC}^{} \equiv \displaystyle G_{F}^{} n_{n}^{}/\sqrt{2}$ ($n_{e}^{}$ and $n_{n}^{}$ are the electron and neutron number densities, respectively) are the CC and the NC contributions to the neutrino matter potentials respectvely. $U$ is just the $3 \times 4$ non-unitary mixing matrix given in Eq.~(6).

Note that in Eq.~(8), $E$ is a $4 \times 4$ diagonal matrix because altogether four mass eigenstates $\nu^{}_{1}$, $\nu^{}_{2}$, $\nu^{}_{3}$ and $\nu^{}_{4}$ can be kinematically produced and therefore participate in the neutrino oscillation, while $\bar{A}$ is a $3 \times 3$ matrix since only three active left-handed neutrinos $\nu^{}_{e}$, $\nu^{}_{\mu}$ and $\nu^{}_{\tau}$ are involved in the CC or the NC interactions. The Hamiltonian in Eq.~(8) holds for neutrinos, whereas one has to perform the replacements $U \rightarrow U^{*}_{}$, $V^{}_{CC} \rightarrow - V^{}_{CC}$ and $V^{}_{NC} \rightarrow - V^{}_{NC}$ for antineutrinos.

In the case of matter of constant density, the Hermitian matrix $\tilde{{\cal H}}$ can be diagonalized by a unitary transformation $\tilde{{\cal H}} = X^{}_{} \tilde{E} X^{\dagger}_{}$, where $\tilde{E} \equiv {\rm diag} \left \{ \tilde{E}^{}_{1}, \tilde{E}^{}_{2}, \tilde{E}^{}_{3}, \tilde{E}^{}_{4} \right \}$ is the effective energy matrix in matter and $X$ is a $4 \times 4$ unitary matrix. It can be inferred that the matrix $\tilde{U} \equiv U X_{}^{*}$, which is also a $3 \times 4$ non-unitary matrix, can be regarded as the effective leptonic mixing matrix in matter. Therefore, we can write down the neutrino oscillation probabilities in matter as
\begin{eqnarray}
\tilde{P} \; ( \stackrel{(-)}{\nu}^{}_{\alpha} \rightarrow \stackrel{(-)}{\nu}^{}_{\beta} ) & = & \frac{1}{ \Big ( \displaystyle \sum^{}_{i=1,2,3,4} | \tilde{U}^{}_{\alpha i} |^2 \Big ) \Big ( \displaystyle \sum^{}_{i=1,2,3,4} | \tilde{U}^{}_{\beta i} |^2 \Big )} \Bigg \{ \left | \sum^{}_{i=1,2,3,4} \tilde{U}^{*}_{\alpha i} \tilde{U}^{}_{\beta i} \right |^2 \nonumber\\
& & - 4 \sum^{}_{j > i} {\rm Re} \left [ \tilde{U}^{}_{\alpha i} \tilde{U}^{}_{\beta j} \tilde{U}^{*}_{\alpha j} \tilde{U}^{*}_{\beta i} \right ] \sin^2 \tilde{\Delta}^{}_{ji} \pm 2 \sum^{}_{j > i} {\rm Im} \left [ \tilde{U}^{}_{\alpha i} \tilde{U}^{}_{\beta j} \tilde{U}^{*}_{\alpha j} \tilde{U}^{*}_{\beta i} \right ] \sin2\tilde{\Delta}^{}_{ji} \Bigg \} \; ,
\end{eqnarray}
where $\tilde{\Delta}^{}_{ji} \equiv \Delta \tilde{m}^2_{ji} L / 4E$ with $\Delta \tilde{m}^2_{ji} \equiv \tilde{m}^{2}_{j} - \tilde{m}^{2}_{i} = 2 E^{}_{\nu} ( \tilde{E}^{}_{j} - \tilde{E}^{}_{i} )$ is the effective neutrino mass-squared difference in matter.

One can find from Eqs.~(7) and (9), the UV in the MNSP matrix may result in three kinds of different effects on the neutrino oscillation probabilities summarised as follows.
\begin{itemize}
\item The IUV can affect the neutrino production and detection, and then generate the ``zero-distance'' effect \cite{zero}, i.e., at $L = 0$ it follows that
\begin{eqnarray}
P \; ( \stackrel{(-)}{\nu}^{}_{\alpha} \rightarrow  \stackrel{(-)}{\nu}^{}_{\beta} ) |_{L = 0} \; = \; \frac{\left | \left ( U U^{\dagger}_{} \right )_{\beta \alpha} \right |^2}{\left ( U U^{\dagger}_{} \right )_{\alpha \alpha} \left ( U U^{\dagger}_{} \right )_{\beta \beta}} \; \neq \; 0 ~~~~~~ (\alpha \neq \beta) \; ,
\end{eqnarray}
where $U$ is the $3 \times 4$ matrix in Eq.~(6). It means the flavor transition already took place at the source before the oscillation begins, with very tiny transition probability of the order $s^{4}_{}$. It is worth to mention that the ``zero-distance'' effect is irrelevant to the DUV, which means this effect would not be observed if there exists only the light sterile neutrinos.
\item The existence of light sterile neutrinos will introduce both additional CP-conserving and CP-violating oscillatory terms in neutrino oscillation probabilities. Their oscillation amplitudes are of the order $s^{2}_{}$ and the frequencies are proportional to the newly introduced mass-squared differences. The oscillatory behaviors of these new frequencies may be observed at very-short baseline neutrino oscillation experiments \cite{sterile}. 
\item In the propagation, both the DUV and the IUV can modify the amplitudes of the standard oscillatory terms as well as affect the matter effects. Moreover, for the neutrino beams with relative hight energies, the non-unitary effects may be largely enhanced by the neutrino-matter interactions \cite{matter}.
\end{itemize}

Although by using Eqs.~(7) - (9), the oscillation probabilities can be numerically calculated with any required accuracy, the analytical approximate formulas can be useful in showing the UV effects in a more transparent way. In the next two sections, we are going to discuss both the analytical expansions and numerical calculations of the neutrino oscillation probabilities in the case of UV.

Indeed, there are several previous publications discussing the hybrid scenarios with both light and heavy sterile neutrinos~\cite{Nelson:2010hz,Kuflik:2012sw}, however, our study is different because the present work presents the long-baseline neutrino oscillation behaviors in the presence of Earth matter effects, while the previous ones focus on the property of short baseline neutrino oscillations and the explanation of experimental anomalies. For the first time, we derive a complete set of series expansion formulas for neutrino oscillation probabilities in matter of constant density in the presence of both light and heavy sterile neutrinos. Formulas for the ($3$+$\mathbbm{1}$+$\mathbf{2}$) scenario can be easily generalized to scenarios with arbitrary numbers of light and heavy sterile neutrinos. Note that sterile neutrino induced CP violation may have significant observable effects in current and future long-baseline neutrino experiments (e.g., T2K~\cite{Klop:2014ima,Palazzo:2015gja}, DUNE~\cite{Berryman:2015nua}), which makes our derivations important and timely to understand these possible distinct phenomena in long-baseline experiments.

\section{Analytical Expansions of Neutrino Oscillation Probabilities}

In this section, we derive the series expansion formulas for neutrino oscillation probabilities in matter of constant density by using the perturbation theory. The formulas are expanded up to the first order of both the mass hierarchy parameter $\alpha \equiv \Delta m_{21}^{2} / \Delta m_{31}^{2}$ and the UV parameters $s^{2}_{ij}\;(i = 1, 2, 3\;{\rm and}\;j = 4, 5, 6)$. The details of the diagonalization of the Hamiltonian  $\tilde{{\cal H}}$ can be found in Appendix A, and the resulting approximate neutrino oscillation probabilities in matter are presented in Appendix B, where $\tilde{P}^{}_{\alpha \beta} \equiv \tilde{P} ( \nu^{}_{\alpha} \rightarrow \nu^{}_{\beta} )$ is the transition probability from a neutrino flavor $\alpha$ to a neutrino flavor $\beta$ in matter.
Equations (B3) - (B8) and Eqs.~(B10) - (B12) can be further simplified if one can neglect those terms of ${\cal O} (s^{2}_{13} \alpha)$, ${\cal O} (s^{2}_{13} s^{2}_{ij})$ or ${\cal O} (s^{4}_{13})$, the approximate neutrino oscillation probabilities can then be obtained as:
\begin{eqnarray}
\tilde{P}^{}_{e e} & \approx & 1 - 2 s^{2}_{14} - \frac{4 s^{}_{13}}{\left ( 1 - A^{}_{CC} \right )^2} \left (s^{}_{13} + 2 R^{}_{b} A^{}_{NC} \right ) \sin^2 (1-A^{}_{CC})\Delta^{}_{31}  \; , \\[3mm]
\tilde{P}^{}_{\mu \mu} & \approx & 1 - 2 s^{2}_{24} - \left [ \sin^2 2\theta^{}_{23} \left ( 1 - 2 s^{2}_{24} \right ) - \frac{ \alpha (1 + A^{}_{CC}) \sin2\theta^{}_{12} \sin4\theta^{}_{23} s^{}_{13} \cos\delta}{A^{}_{CC}} \right. \nonumber\\[3mm]
& & \left. + 2 \sin4\theta^{}_{23} \left ( \frac{R^{}_{a} A^{}_{NC} s^{}_{13}}{A^{}_{CC}} + T^{}_{a} A^{}_{NC} \right ) \right ] \sin^2 \Delta^{}_{31} \nonumber\\[3mm]
& & + \frac{4 s^{}_{23} s^{}_{13}}{1 - A^{}_{CC}} \left ( \frac{s^{}_{23} s^{}_{13}}{1 - A^{}_{CC}} - \frac{\alpha \sin2\theta^{}_{12} c^{}_{23} \cos\delta}{A^{}_{CC}} + 2 R^{}_{12} + \frac{2 R^{}_{a} A^{}_{NC} c^{}_{23}}{A^{}_{CC}} + \frac{2 R^{}_{b} A^{}_{NC} s^{}_{23}}{1 - A^{}_{CC}} \right ) \nonumber\\[3mm]
& & \cdot \left [ c^{2}_{23} \sin(1+A^{}_{CC})\Delta^{}_{31} - s^{2}_{23} \sin(1-A^{}_{CC})\Delta^{}_{31} \right ] \sin(1-A^{}_{CC})\Delta^{}_{31} \nonumber\\[3mm]
& & + \Delta^{}_{31} \sin^2 2\theta^{}_{23} \left ( \alpha c^{2}_{12} - \frac{A^{}_{CC} s^{2}_{13}}{1 - A^{}_{CC}} - \frac{R^{}_{b} A^{}_{NC} s^{}_{13}}{1 - A^{}_{CC}} - \frac{1}{2} T^{}_{c} A^{}_{NC} \right ) \sin2\Delta^{}_{31} \; , \\[3mm]
\tilde{P}^{}_{\tau \tau} & \approx & 1 -  2 s^{2}_{34} - \left [ \sin^2 2\theta^{}_{23} \left ( 1 - 2 s^{2}_{34} \right ) - \frac{ \alpha (1 + A^{}_{CC}) \sin2\theta^{}_{12} \sin4\theta^{}_{23} s^{}_{13} \cos\delta}{A^{}_{CC}} \right. \nonumber\\[3mm]
& & \left. + 2 \sin4\theta^{}_{23} \left ( R^{}_{23} + \frac{R^{}_{a} A^{}_{NC} s^{}_{13}}{A^{}_{CC}} + T^{}_{a} A^{}_{NC} \right ) \right ] \sin^2 \Delta^{}_{31} \nonumber\\[3mm]
& & + \frac{4 c^{}_{23} s^{}_{13}}{1 - A^{}_{CC}} \left ( \frac{c^{}_{23} s^{}_{13}}{1 - A^{}_{CC}} +\frac{\alpha \sin2\theta^{}_{12} s^{}_{23} \cos\delta}{A^{}_{CC}} + 2 R^{}_{13} - \frac{2 R^{}_{a} A^{}_{NC} s^{}_{23}}{A^{}_{CC}} + \frac{2 R^{}_{b} A^{}_{NC} c^{}_{23}}{1 - A^{}_{CC}} \right ) \nonumber\\[3mm]
& & \cdot \left [ s^{2}_{23} \sin(1+A^{}_{CC})\Delta^{}_{31} - c^{2}_{23} \sin(1-A^{}_{CC})\Delta^{}_{31} \right ] \sin(1-A^{}_{CC})\Delta^{}_{31} \nonumber\\[3mm]
& & + \Delta^{}_{31} \sin^2 2\theta^{}_{23} \left ( \alpha c^{2}_{12} - \frac{A^{}_{CC} s^{2}_{13}}{1 - A^{}_{CC}} - \frac{R^{}_{b} A^{}_{NC} s^{}_{13}}{1 - A^{}_{CC}} - \frac{1}{2} T^{}_{c} A^{}_{NC} \right ) \sin2\Delta^{}_{31} \; , \\[3mm]
\tilde{P}^{}_{e \mu} & \approx & \frac{4 s^{}_{23} s^{}_{13}}{1 - A^{}_{CC}} \left \{ \frac{s^{}_{23}}{1 - A^{}_{CC}} \left ( s^{}_{13} + 2 R^{}_{b} A^{}_{NC} \right ) \sin(1-A^{}_{CC})\Delta^{}_{31} \right. \nonumber\\[3mm]
& & + R^{}_{12} \sin(1-A^{}_{CC})\Delta^{}_{31} + I^{}_{12} \cos(1-A^{}_{CC})\Delta^{}_{31} \nonumber\\[3mm]
& & + \frac{c^{}_{23}}{A^{}_{CC}} \left [ \left ( \alpha \sin2\theta^{}_{12} \cos\delta - 2 R^{}_{a} A^{}_{NC} \right ) \cos\Delta^{}_{31} \right. \nonumber\\[3mm]
& & \left. \left. + \left ( \alpha \sin2\theta^{}_{12} \sin\delta + 2 I^{}_{a} A^{}_{NC} \right ) \sin\Delta^{}_{31} \right ] \sin(A^{}_{CC}\Delta^{}_{31}) \right \} \sin(1-A^{}_{CC})\Delta^{}_{31} \; , \\[3mm]
\tilde{P}^{}_{e \tau} & \approx & \frac{4 c^{}_{23} s^{}_{13}}{1 - A^{}_{CC}} \left \{ \frac{c^{}_{23}}{1 - A^{}_{CC}} \left ( s^{}_{13} + 2 R^{}_{b} A^{}_{NC} \right ) \sin(1-A^{}_{CC})\Delta^{}_{31} \right. \nonumber\\[3mm]
& & + R^{}_{13} \sin(1-A^{}_{CC})\Delta^{}_{31} + I^{}_{13} \cos(1-A^{}_{CC})\Delta^{}_{31} \nonumber\\[3mm]
& & - \frac{s^{}_{23}}{A^{}_{CC}} \left [ \left ( \alpha \sin2\theta^{}_{12} \cos\delta - 2 R^{}_{a} A^{}_{NC} \right ) \cos\Delta^{}_{31} \right. \nonumber\\[3mm]
& & \left. \left. + \left ( \alpha \sin2\theta^{}_{12} \sin\delta + 2 I^{}_{a} A^{}_{NC} \right ) \sin\Delta^{}_{31} \right ] \sin(A^{}_{CC}\Delta^{}_{31}) \right \} \sin(1-A^{}_{CC})\Delta^{}_{31} \; , \\[3mm]
\tilde{P}^{}_{\mu \tau} & \approx & \left [ \sin^2 2\theta^{}_{23} \left ( 1 - s^{2}_{24} - s^{2}_{34} \right ) - \frac{\alpha \left ( 1 + A^{}_{CC} \right ) \sin2\theta^{}_{12} \sin4\theta^{}_{23} s^{}_{13} \cos\delta}{A^{}_{CC}} \right. \nonumber\\[3mm]
& & \left. + 2 \sin4\theta^{}_{23} \left ( R^{}_{23} + \frac{R^{}_{a} A^{}_{NC} s^{}_{13}}{A^{}_{CC}} + T^{}_{a} A^{}_{NC} \right ) \right ] \sin^2 \Delta^{}_{31} \nonumber\\[3mm]
& & - \frac{\sin^2 2\theta^{}_{23} s^{2}_{13}}{\left ( 1 - A^{}_{CC} \right )^2} \sin^2 (1-A^{}_{CC})\Delta^{}_{31} + I^{}_{23} \sin2\theta^{}_{23} \sin2\Delta^{}_{31} \nonumber\\[3mm]
& & + \Delta^{}_{31} \sin^2 2\theta^{}_{23} \left [ - \alpha c^{2}_{12} + \frac{A^{}_{CC} s^{2}_{13}}{1 - A^{}_{CC}} + \frac{R^{}_{b} A^{}_{NC} s^{}_{13}}{1 - A^{}_{CC}} + T^{}_{c} A^{}_{NC} \right ] \sin2\Delta^{}_{31} \nonumber\\[3mm]
& & + \frac{2 \sin2\theta^{}_{23} s^{}_{13}}{1 - A^{}_{CC}} \left [ \left ( \frac{\alpha \sin2\theta^{}_{12} \cos2\theta^{}_{23} \cos\delta}{A^{}_{CC}} - 2 R^{}_{c} \right ) \cos(A^{}_{CC}\Delta^{}_{31}) \right. \nonumber\\[3mm]
& & \left. + \left ( \frac{\alpha \sin2\theta^{}_{12} \sin\delta}{A^{}_{CC}} - 2 I^{}_{a} \right ) \sin(A^{}_{CC}\Delta^{}_{31}) \right ] \sin\Delta^{}_{31} \sin(1-A^{}_{CC})\Delta^{}_{31} \nonumber\\[3mm]
& & - 4 A^{}_{NC} \sin2\theta^{}_{23} s^{}_{13} \left [ \frac{1}{1 - A^{}_{CC}} \left ( \frac{R^{}_{a} \cos2\theta^{}_{23}}{A^{}_{CC}} + \frac{R^{}_{b} \sin2\theta^{}_{23}}{1 - A^{}_{CC}} \right ) \cos(A^{}_{CC}\Delta^{}_{31}) \right. \nonumber\\[3mm]
& & \left. - \frac{I^{}_{a}}{A^{}_{CC}} \sin(A^{}_{CC}\Delta^{}_{31}) \right ] \sin\Delta^{}_{31} \sin(1-A^{}_{CC})\Delta^{}_{31} \; .
\end{eqnarray}
Here the expressions of $R^{}_{12}$, $R^{}_{13}$, $R^{}_{23}$, $I^{}_{12}$, $I^{}_{13}$, $I^{}_{23}$, $R^{}_{a}$, $R^{}_{b}$, $R^{}_{c}$, $I^{}_{a}$, $T^{}_{a}$ and $T^{}_{c}$ can be found in Eqs.~(A2) and (A3).
Due to the existence of UV (both the direct and the indirect), the ``to all'' transition probability of the $\alpha$ flavor to all active flavors $\tilde{P}(\nu^{}_{\alpha} \rightarrow \nu^{}_{e, \mu, \tau}) \equiv \tilde{P}^{}_{\alpha e} + \tilde{P}^{}_{\alpha \mu} +\tilde{P}^{}_{\alpha \tau}$ (for $\alpha = e, \mu, \tau$) is not the unity:
\begin{eqnarray}
\tilde{P}(\nu^{}_{e} \rightarrow \nu^{}_{e, \mu, \tau}) & \approx & 1 - 2 s^{2}_{14} + \frac{4 s^{}_{13}}{1 - A^{}_{CC}} \left [ R^{}_{b} \sin(1-A^{}_{CC})\Delta^{}_{31} \right. \nonumber\\[3mm]
& & \left. + I^{}_{b} \cos(1-A^{}_{CC})\Delta^{}_{31} \right ] \sin(1-A^{}_{CC})\Delta^{}_{31} \; , \\[3mm]
\tilde{P}(\nu^{}_{\mu} \rightarrow \nu^{}_{e, \mu, \tau}) & \approx & 1 - 2 s^{2}_{24} + \left [ \sin^2 2\theta^{}_{23} \left ( s^{2}_{24} - s^{2}_{34} \right ) + R^{}_{23} \sin4\theta^{}_{23} \right ] \sin^2 \Delta^{}_{31} + I^{}_{23} \sin2\theta^{}_{23} \sin2\Delta^{}_{31} \nonumber\\[3mm]
& & + \frac{4 s^{}_{23} s^{}_{13}}{1 - A^{}_{CC}} \left \{ R^{}_{12} \left [ c^{2}_{23} \sin(1+A^{}_{CC})\Delta^{}_{31} - s^{2}_{23} \sin(1-A^{}_{CC})\Delta^{}_{31} \right ] \right. \nonumber\\[3mm]
& & - I^{}_{12} \left [ c^{2}_{23} \cos(1+A^{}_{CC})\Delta^{}_{31} + s^{2}_{23} \cos(1-A^{}_{CC})\Delta^{}_{31} \right ] \nonumber\\[3mm]
& & \left. - \sin2\theta^{}_{23} \left [ R^{}_{13} \cos(A^{}_{CC}\Delta^{}_{31}) + I^{}_{13} \sin(A^{}_{CC}\Delta^{}_{31}) \right ] \sin\Delta^{}_{31} \right \} \sin(1-A^{}_{CC})\Delta^{}_{31} , \\[3mm]
\tilde{P}(\nu^{}_{\tau} \rightarrow \nu^{}_{e, \mu, \tau}) & \approx & 1 - 2 s^{2}_{34} - \left [ \sin^2 2\theta^{}_{23} \left ( s^{2}_{24} - s^{2}_{34} \right ) + R^{}_{23} \sin4\theta^{}_{23} \right ] \sin^2 \Delta^{}_{31} - I^{}_{23} \sin2\theta^{}_{23} \sin2\Delta^{}_{31} \nonumber\\[3mm]
& & - \frac{4 c^{}_{23} s^{}_{13}}{1 - A^{}_{CC}} \left \{ \sin2\theta^{}_{23} \left [ R^{}_{12} \cos(A^{}_{CC}\Delta^{}_{31}) + I^{}_{12} \sin(A^{}_{CC}\Delta^{}_{31}) \right ] \sin\Delta^{}_{31} \right. \nonumber\\[3mm]
& & - R^{}_{13} \left [ s^{2}_{23} \sin(1+A^{}_{CC})\Delta^{}_{31} - c^{2}_{23} \sin(1-A^{}_{CC})\Delta^{}_{31} \right ] \nonumber\\[3mm]
& & \left. + I^{}_{13} \left [ s^{2}_{23} \cos(1+A^{}_{CC})\Delta^{}_{31} + c^{2}_{23} \cos(1-A^{}_{CC})\Delta^{}_{31} \right ] \right \} \sin(1-A^{}_{CC})\Delta^{}_{31} \; .
\end{eqnarray}
It is interesting to find that these summed probabilities are all independent of the NC potential $V^{}_{NC}$ in these first-order expansions.

Up to the same order, the corresponding probabilities in vacuum ${P}^{}_{\alpha \beta}$ and $P(\nu^{}_{\alpha} \rightarrow \nu^{}_{e, \mu, \tau})$ in the case of UV are given for comparison:
\begin{eqnarray}
P^{}_{e e} & \approx & 1 - 2 s^{2}_{14} - \sin^2 2\theta^{}_{13} \sin^2 \Delta^{}_{31}  \; , \\[3mm]
P^{}_{\mu \mu} & \approx & 1 - 2 s^{2}_{24} - \left [ \left ( 1 - 2 s^{2}_{24} \right ) \sin^2 2\theta^{}_{23} - 4\cos2\theta^{}_{23} s^{2}_{23} s^{2}_{13} - 8 R^{}_{12} \cos2\theta^{}_{23} s^{}_{23} s^{}_{13} \right ] \sin^2 \Delta^{}_{31} \nonumber\\[3mm]
& & + \alpha \Delta^{}_{31} \sin2\theta^{}_{23} \left ( c^{2}_{12} \sin2\theta^{}_{23} - 2 \sin2\theta^{}_{12} s^{2}_{23} s^{}_{13} \cos\delta \right ) \sin2\Delta^{}_{31} \; , \\[3mm]
P^{}_{\tau \tau} & \approx & 1 - 2 s^{2}_{34} - \left [ \left ( 1 - 2 s^{2}_{34} \right ) \sin^2 2\theta^{}_{23} + 4\cos2\theta^{}_{23} c^{2}_{23} s^{2}_{13} \right. \nonumber\\[3mm]
& & \left. + 8 \left ( R^{}_{12} s^{}_{13} + R^{}_{23} s^{}_{23} \right ) \cos2\theta^{}_{23} c^{}_{23} \right ] \sin^2 \Delta^{}_{31} \nonumber\\[3mm]
& & + \alpha \Delta^{}_{31} \sin2\theta^{}_{23} \left ( c^{2}_{12} \sin2\theta^{}_{23} + 2 \sin2\theta^{}_{12} c^{2}_{23} s^{}_{13} \cos\delta \right ) \sin2\Delta^{}_{31} \; , \\[3mm]
P^{}_{e \mu} & \approx & \left [ s^{2}_{23} \sin^2 2\theta^{}_{13} + 4 R^{}_{12} s^{}_{23} s^{}_{13} + 2 \alpha \Delta^{}_{31} \sin2\theta^{}_{12} \sin2\theta^{}_{23} s^{}_{13} \sin\delta \right ] \sin^2 \Delta^{}_{31} \nonumber\\[3mm]
& & + \left ( 2 I^{}_{12} s^{}_{23} s^{}_{13} + \alpha \Delta^{}_{31} \sin2\theta^{}_{12} \sin2\theta^{}_{23} s^{}_{13} \cos\delta \right ) \sin2\Delta^{}_{31} \; , \\[3mm]
P^{}_{e \tau} & \approx & \left [ c^{2}_{23} \sin^2 2\theta^{}_{13} + 4 R^{}_{13} c^{}_{23} s^{}_{13} - 2 \alpha \Delta^{}_{31} \sin2\theta^{}_{12} \sin2\theta^{}_{23} s^{}_{13} \sin\delta \right ] \sin^2 \Delta^{}_{31} \nonumber\\[3mm]
& & + \left ( 2 I^{}_{13} c^{}_{23} s^{}_{13} - \alpha \Delta^{}_{31} \sin2\theta^{}_{12} \sin2\theta^{}_{23} s^{}_{13} \cos\delta \right ) \sin2\Delta^{}_{31} \; , \\[3mm]
P^{}_{\mu \tau} & \approx & \sin2\theta^{}_{23} \left \{ \left [ \sin2\theta^{}_{23} c^{4}_{13} - \left ( s^{2}_{24} + s^{2}_{34} \right ) \sin2\theta^{}_{23} - 4 R^{}_{c} s^{}_{13} \right. \right.  \nonumber\\[3mm]
& & \left. + 2 R^{}_{23} \cos2\theta^{}_{23} + 2 \alpha \Delta^{}_{31} \sin2\theta^{}_{12} s^{}_{13} \sin\delta \right ] \sin^2 \Delta^{}_{31} \nonumber\\[3mm]
& & \left. + \left [ I^{}_{23} - \alpha \Delta^{}_{31} \left ( c^{2}_{12} \sin2\theta^{}_{23} + \sin2\theta^{}_{12} \cos2\theta^{}_{23} s^{}_{13} \cos\delta \right ) \right ] \sin2\Delta^{}_{31} \right \} \; .
\end{eqnarray}
And
\begin{eqnarray}
P(\nu^{}_{e} \rightarrow \nu^{}_{e, \mu, \tau}) & \approx & 1 - 2 s^{2}_{14} + 4 R^{}_{b} s^{}_{13} \sin^2 \Delta^{}_{31} + 2 I^{}_{b} s^{}_{13} \sin2\Delta^{}_{31} \; , \\[3mm]
P(\nu^{}_{\mu} \rightarrow \nu^{}_{e, \mu, \tau}) & \approx & 1 - 2 s^{2}_{24} + \left [ \left ( s^{2}_{24} - s^{2}_{34} \right ) \sin^2 2\theta^{}_{23} + 4 \left ( R^{}_{12} \cos2\theta^{}_{23} - R^{}_{13} \sin2\theta^{}_{23} \right ) s^{}_{23} s^{}_{13} \right. \nonumber\\[3mm]
& & \left. + R^{}_{23} \sin4\theta^{}_{23} \right ] \sin^2 \Delta^{}_{31} + \left( I^{}_{23} \sin2\theta^{}_{23} - 2 I^{}_{12} s^{}_{23} s^{}_{13} \right ) \sin2\Delta^{}_{31} \; , \\[3mm]
P(\nu^{}_{\tau} \rightarrow \nu^{}_{e, \mu, \tau}) & \approx & 1 - 2 s^{2}_{34} - \left [ \left ( s^{2}_{24} - s^{2}_{34} \right ) \sin^2 2\theta^{}_{23} + 4 \left ( R^{}_{12} \sin2\theta^{}_{23} + R^{}_{13} \cos2\theta^{}_{23} \right ) c^{}_{23} s^{}_{13} \right. \nonumber\\[3mm]
& & \left. + R^{}_{23} \sin4\theta^{}_{23} \right ] \sin^2 \Delta^{}_{31} - \left( I^{}_{23} \sin2\theta^{}_{23} + 2 I^{}_{13} c^{}_{23} s^{}_{13} \right ) \sin2\Delta^{}_{31} \; .
\end{eqnarray}
The approximate oscillation probabilities in the standard three-flavor case can be easily obtained by switching off the active-sterile mixing (i.e., setting all $\theta^{}_{ij} = 0$, where $i = 1, 2, 3$ and $j = 4, 5, 6$). In this limit, our results are consistent with those obtained in Ref.~\cite{Arafune:1997hd,Cervera:2000kp,Freund:2001pn,Akhmedov2004,Xu:2015kma}.

\section{Numerical calculations and Discussion}

\begin{figure}
\begin{center}
\vspace{0cm}
\includegraphics[scale=0.75, angle=0, clip=0]{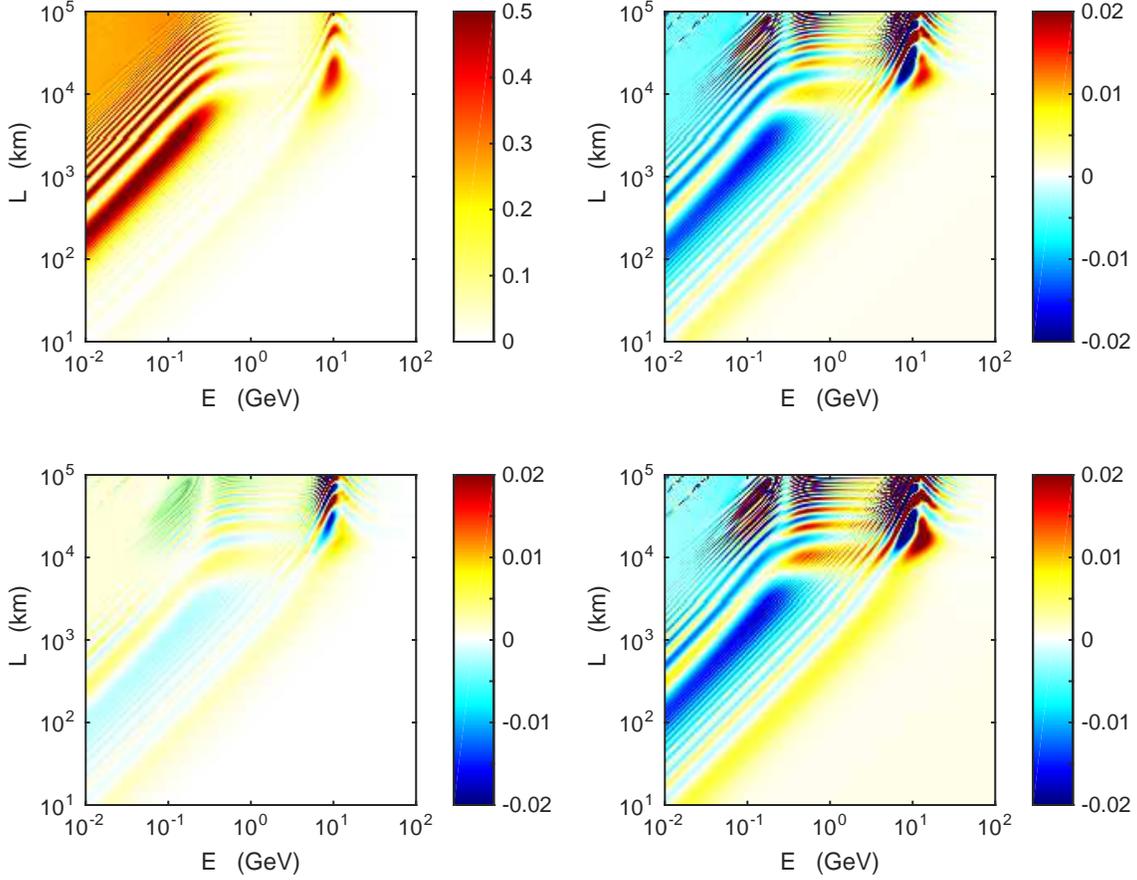}
\vspace{-0.3cm}
\caption{Contours of the probability $\tilde{P}^{}_{e \mu}$ calculated numerically without any approximations in the standard (3+0+0) scenario (top-left) and its corrections by different UV effects $\tilde{P}^{\rm UV}_{e \mu} - \tilde{P}^{\rm standard}_{e \mu}$ in the (3+0+2) (bottom-left), (3+1+0) (top-right) and (3+1+2) (bottom-right) scenarios as functions of the neutrino energy $E$ and the baseline length $L$. The probabilities are averaged over a Gaussian energy resolution of $2\%$.}
\end{center}
\end{figure}

\begin{figure}
\begin{center}
\vspace{-0.4cm}
\includegraphics[scale=0.7, angle=0, clip=0]{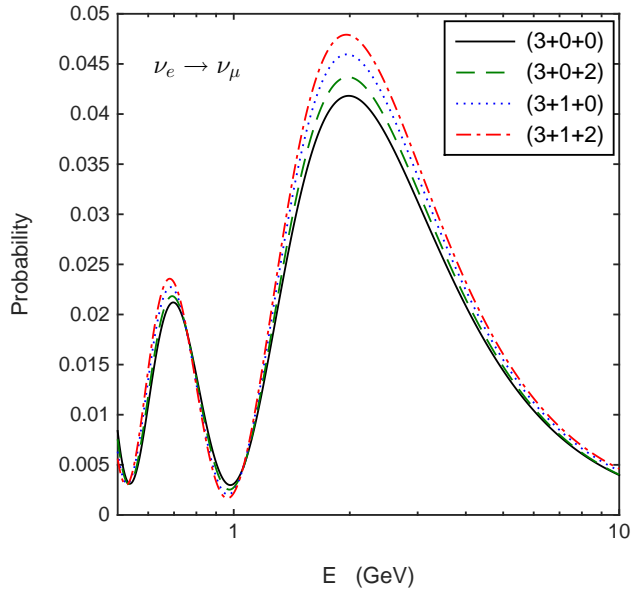}
\vspace{-0.4cm}
\caption{The probabilities $\tilde{P}^{}_{e \mu}$ at the baseline $L = 1000$ km calculated numerically without any approximations as functions of the neutrino energy $E$ in the four different scenarios. The probabilities are averaged over a Gaussian energy resolution of $2\%$.}
\end{center}
\end{figure}

\begin{figure}
\begin{center}
\vspace{-0.4cm}
\includegraphics[scale=0.7, angle=0, clip=0]{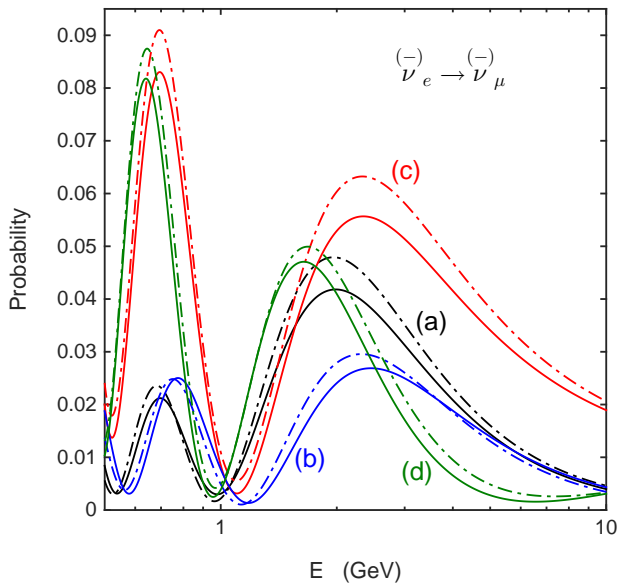}
\vspace{-0.4cm}
\caption{The probabilities $\tilde{P}^{}_{e \mu}$ ($\tilde{P}^{}_{\bar{e} \bar{\mu}}$) at the baseline $L = 1000$ km calculated numerically without any approximations as functions of the neutrino energy $E$ in the (3+0+0) scenario (solid lines) and the (3+1+2) scenario (dash-dotted lines) for (a) $\nu^{}_{e} \rightarrow \nu^{}_{\mu}$ oscillation in the NH case, (b) $\nu^{}_{e} \rightarrow \nu^{}_{\mu}$ oscillation in the IH case, (c) $\bar{\nu}^{}_{e} \rightarrow \bar{\nu}^{}_{\mu}$ oscillation in the NH case and (d) $\bar{\nu}^{}_{e} \rightarrow \bar{\nu}^{}_{\mu}$ oscillation in the IH case. The probabilities are averaged over a Gaussian energy resolution of $2\%$.}
\end{center}
\end{figure}

\begin{figure}
\begin{center}
\vspace{-0.6cm}
\includegraphics[scale=0.7, angle=0, clip=0]{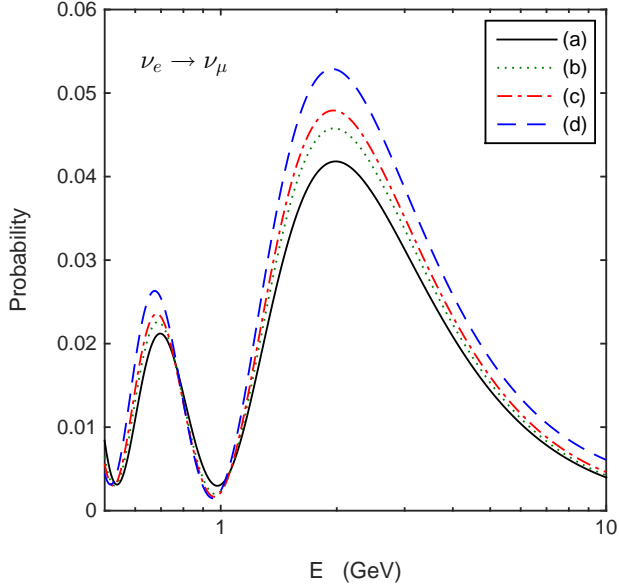}
\vspace{-0.5cm}
\caption{The probabilities $\tilde{P}^{}_{e \mu}$ at the baseline $L = 1000$ km calculated numerically without any approximations as functions of the neutrino energy $E$ in the (3+0+0) scenario (case (a)) and the (3+1+2) scenario with $\theta^{}_{14} = \theta^{}_{24} = \theta^{}_{34} = 5^{\circ}$ (case (b)), $7^{\circ}$ (case (c)) or $10^{\circ}$ (case (d)). The probabilities are averaged over a Gaussian energy resolution of $2\%$.}
\end{center}
\end{figure}

\begin{figure}
\begin{center}
\vspace{-0.6cm}
\includegraphics[scale=0.7, angle=0, clip=0]{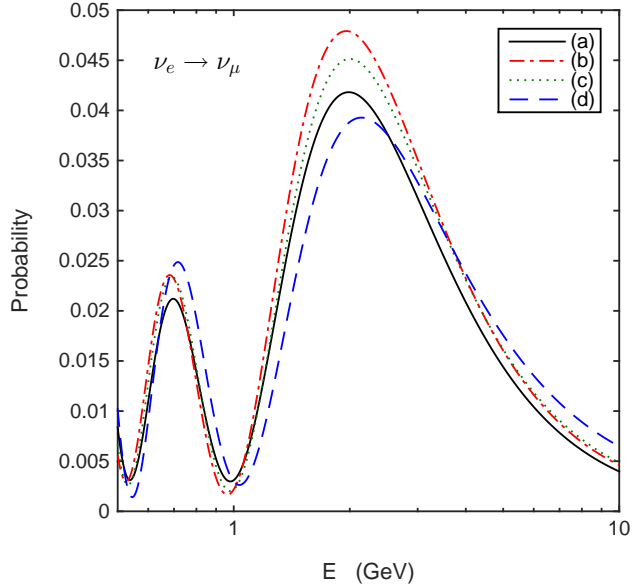}
\vspace{-0.5cm}
\caption{The probabilities $\tilde{P}^{}_{e \mu}$ at the baseline $L = 1000$ km calculated numerically without any approximations as functions of the neutrino energy $E$ in the (3+0+0) scenario (case (a)) and the (3+1+2) scenario with $\delta^{}_{14, 15, 16} = 0$, $\delta^{}_{24, 25, 26} = 60^{\circ}$ and $\delta^{}_{34, 35, 36} = 120^{\circ}$ (case (b)); $\delta^{}_{14, 15, 16} = 0$, $\delta^{}_{24} = 60^{\circ}$, $\delta^{}_{25, 26} = -60^{\circ}$, $\delta^{}_{34} = 120^{\circ}$ and $\delta^{}_{35, 36} = -120^{\circ}$ (case (c)); or $\delta^{}_{14, 15, 16} = 0$, $\delta^{}_{24, 25, 26} = -60^{\circ}$ and $\theta^{}_{34, 35, 36} = -120^{\circ}$ (case (d)). The probabilities are averaged over a Gaussian energy resolution of $2\%$.}
\end{center}
\end{figure}

\begin{figure}
\begin{center}
\vspace{0cm}
\includegraphics[scale=0.75, angle=0, clip=0]{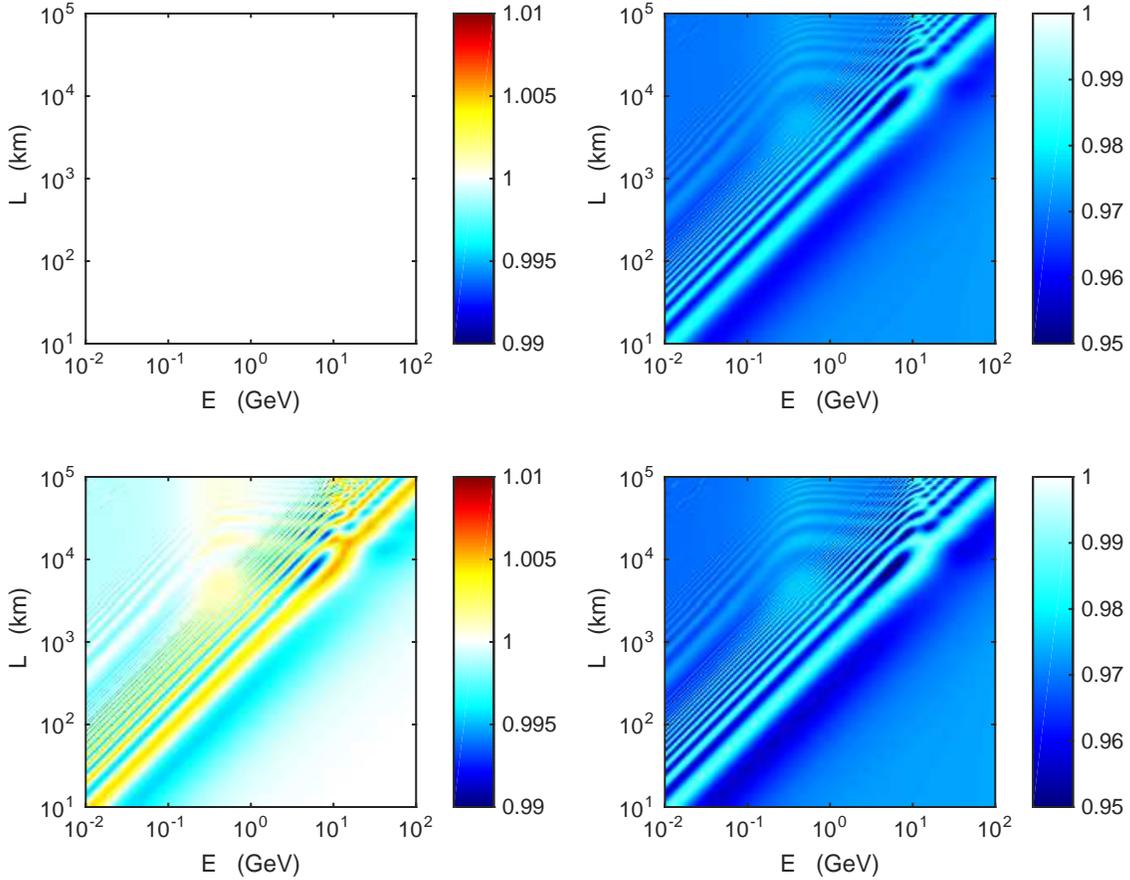}
\vspace{-0.3cm}
\caption{Contours of the probabilities $\tilde{P}(\nu^{}_{\mu} \rightarrow \nu^{}_{e, \mu, \tau})$ calculated numerically without any approximations as functions of the neutrino energy $E$ and the baseline length $L$ in the (3+0+0) (top-left), (3+0+2) (bottom-left), (3+1+0) (top-right) and (3+1+2) (bottom-right) scenarios. The probabilities are averaged over a Gaussian energy resolution of $2\%$.}
\end{center}
\end{figure}

\begin{figure}
\begin{center}
\vspace{-0.4cm}
\includegraphics[scale=0.7, angle=0, clip=0]{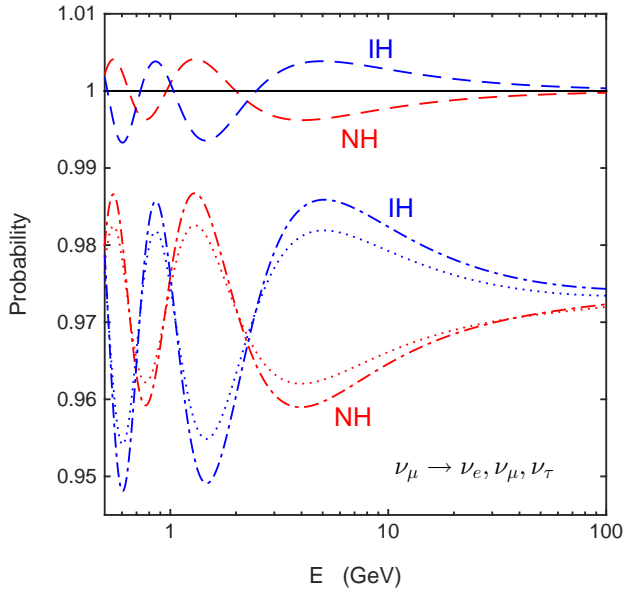}
\vspace{-0.4cm}
\caption{The probabilities $\tilde{P}(\nu^{}_{\mu} \rightarrow \nu^{}_{e, \mu, \tau})$ at the baseline $L = 1000$ km calculated numerically without any approximations as functions of the neutrino energy $E$ in the (3+0+0) (solid lines), (3+0+2) (dashed lines), (3+1+0) (dotted lines) and (3+1+2) (dash-dotted lines) scenarios for both the NH and the IH cases. The probabilities are averaged over a Gaussian energy resolution of $2\%$.}
\end{center}
\end{figure}

\begin{figure}
\begin{center}
\vspace{-0.4cm}
\includegraphics[scale=0.7, angle=0, clip=0]{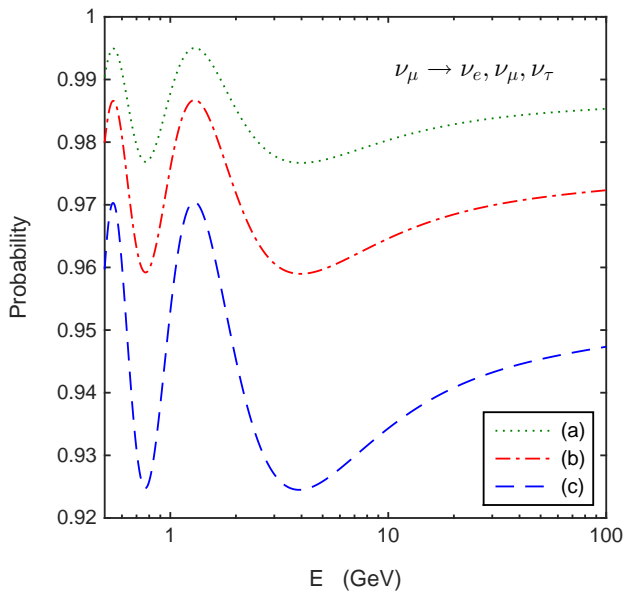}
\vspace{-0.4cm}
\caption{The probabilities $\tilde{P}(\nu^{}_{\mu} \rightarrow \nu^{}_{e, \mu, \tau})$ at the baseline $L = 1000$ km calculated numerically without any approximations as functions of the neutrino energy $E$ in the (3+1+2) scenario with $\theta^{}_{14} = \theta^{}_{24} = \theta^{}_{34} = 5^{\circ}$ (case (a)), $7^{\circ}$ (case (b)) or $10^{\circ}$ (case (c)). The probabilities are averaged over a Gaussian energy resolution of $2\%$.}
\end{center}
\end{figure}

\begin{figure}
\begin{center}
\vspace{0cm}
\includegraphics[scale=0.7, angle=0, clip=0]{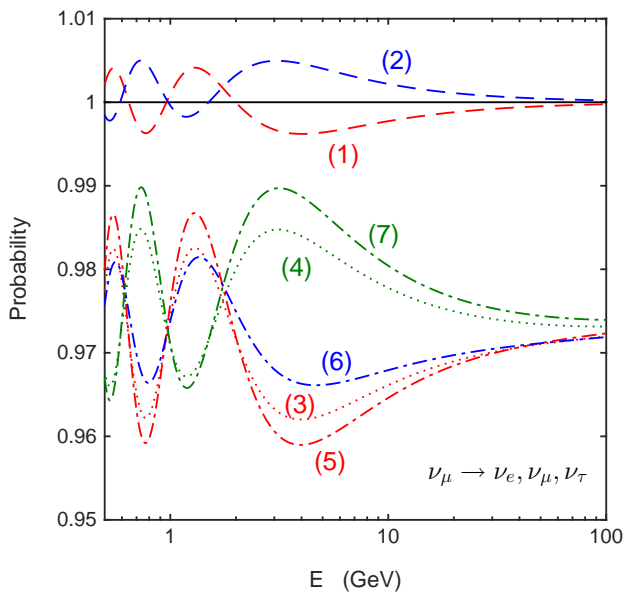}
\vspace{-0.3cm}
\caption{The probabilities $\tilde{P}(\nu^{}_{\mu} \rightarrow \nu^{}_{e, \mu, \tau})$ at the baseline $L = 1000$ km calculated numerically without any approximations as functions of the neutrino energy $E$ in the (3+0+2) scenario with $\delta^{}_{15, 16} = 0$, $\delta^{}_{25, 26} = 60^{\circ}$ and $\delta^{}_{35, 36} = 120^{\circ}$ (case (1)); or $\delta^{}_{15, 16} = 0$, $\delta^{}_{25, 26} = -60^{\circ}$ and $\delta^{}_{35, 36} = -120^{\circ}$ (case (2)); the (3+1+0) scenario with $\delta^{}_{14} = 0$, $\delta^{}_{24} = 60^{\circ}$ and $\delta^{}_{34} = 120^{\circ}$ (case (3)); or $\delta^{}_{14} = 0$, $\delta^{}_{24} = -60^{\circ}$ and $\delta^{}_{34} = -120^{\circ}$ (case (4)); and the (3+1+2) scenario with $\delta^{}_{14, 15, 16} = 0$, $\delta^{}_{24, 25, 26} = 60^{\circ}$ and $\delta^{}_{34, 35, 36} = 120^{\circ}$ (case (5)); $\delta^{}_{14, 15, 16} = 0$, $\delta^{}_{24} = 60^{\circ}$, $\delta^{}_{25, 26} = -60^{\circ}$, $\delta^{}_{34} = 120^{\circ}$ and $\delta^{}_{35, 36} = -120^{\circ}$ (case (6)); or $\delta^{}_{14, 15, 16} = 0$, $\delta^{}_{24, 25, 26} = -60^{\circ}$ and $\theta^{}_{34, 35, 36} = -120^{\circ}$ (case (7)). The probabilities are averaged over a Gaussian energy resolution of $2\%$.}
\end{center}
\end{figure}

\begin{figure}
\begin{center}
\vspace{0cm}
\includegraphics[scale=0.64, angle=0, clip=0]{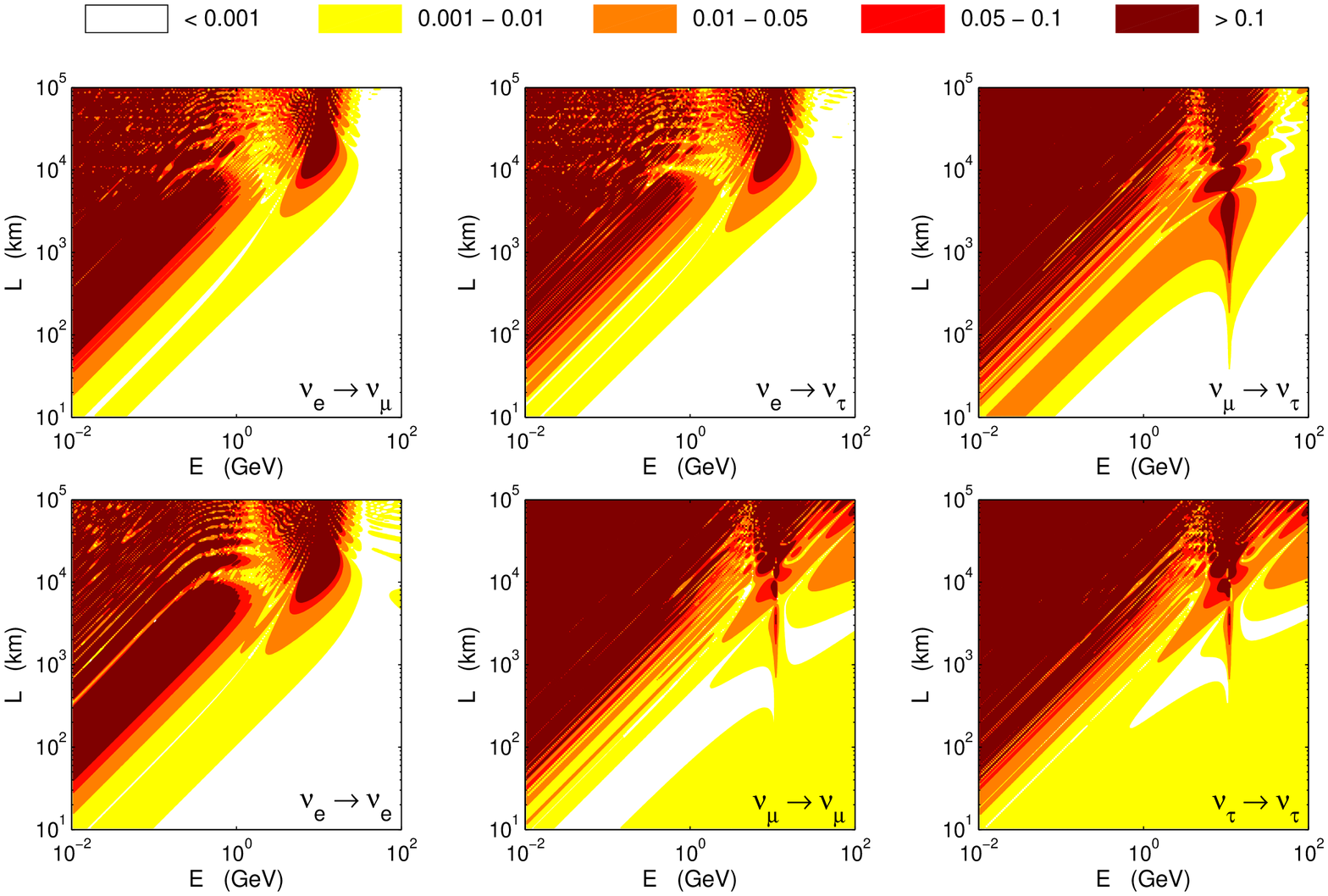}
\vspace{-0.3cm}
\caption{Contours of the absolute errors of the analytical formulas of $\tilde{P}^{}_{\alpha \beta}$ in Eqs.~(11) - (16) as functions of the neutrino energy $E$ and the baseline length $L$.}
\end{center}
\end{figure}

In this section, we carried out the numerical calculations to discuss: i) the accuracy of above approximate formulas and ii) effects of the DUV and the IUV. Here matter of constant density $3\,{\rm g}/{\rm cm}^3$ is assumed and the electron fraction is assumed to be 0.5 \cite{earth}. In our calculation, $\Delta m^{2}_{21} = 7.50 \times 10^{-5}~{\rm eV}^2$, $\Delta m^{2}_{32} = 2.457 \times 10^{-3}~{\rm eV}^2$, $\theta^{}_{12} = 33.48^\circ$, $\theta^{}_{13} = 8.50^\circ$, $\theta^{}_{23} = 42.3^\circ$ and $\delta = 306^\circ$ (or $\Delta m^{2}_{21} = 7.50 \times 10^{-5}~{\rm eV}^2$, $\Delta m^{2}_{32} = 2.449 \times 10^{-3}~{\rm eV}^2$, $\theta^{}_{12} = 33.48^\circ$, $\theta^{}_{13} = 8.51^\circ$, $\theta^{}_{23} = 49.5^\circ$ and $\delta = 254^\circ$) have been taken for the normal hierarchy (NH) case (or the inverted hierarchy (IH) case) \cite{GlobalFit}. For those UV parameters, $\theta^{}_{14} = \theta^{}_{24} = \theta_{34} = 7^\circ$, $\theta^{}_{15} = \theta^{}_{25} = \theta^{}_{35} = \theta^{}_{16} = \theta^{}_{26} = \theta^{}_{36} = 3^\circ$, $\delta^{}_{14} = \delta^{}_{15} = \delta^{}_{16} = 0^\circ$, $\delta^{}_{24} = \delta^{}_{25} = \delta^{}_{26} = 60^\circ$ $\delta^{}_{34} = \delta^{}_{35} = \delta^{}_{36} = 120^\circ$ have been chosen as the default values unless otherwise specified. The newly introduced mass-squared difference $\Delta m^{2}_{41}$ is assumed to be $1.0 ~ {\rm eV}^2$, while the masses of two heavy sterile neutrinos are irrelevant to the calculations. 
This particular choice of the UV parameters in our numerical illustration are motivated from two aspects. On the one hand, the anomalies in short baseline neutrino oscillation experiments suggest that there may exist one or more light sterile neutrinos with the active-sterile mixing of $\cal{O}$(0.1)~\cite{sterile,stefit}. On the other hand, the mixing of heavy sterile neutrinos with the light ones are strongly constrained by the electroweak precision data to be less than $\cal{O}$(0.01) Refs.~\cite{Antusch2006,Antusch2014}. As one can easily image, the larger the active-sterile mixing is, the more significant unitarity violation effects there would be. This is why in our (3+1+2) scenario, the DUV effects dominate over the IUV effects.

Taking the transition probability $\tilde{P}^{}_{e \mu}$ as an example, we illustrate in Fig.~1 the contour structures of this transition probability in the standard (3+0+0) scenario (top-left), and its corrections by different UV effects (i.e., $\tilde{P}^{\rm UV}_{e \mu} - \tilde{P}^{\rm standard}_{e \mu}$) in the (3+0+2) (bottom-left), (3+1+0) (top-right) and (3+1+2) (bottom-right) scenarios as functions of the neutrino energy $E$ and the baseline length $L$. All the probabilities are calculated numerically without any approximations and are averaged over a Gaussian energy resolution of $2\%$. Figure 2 shows how the probability $\tilde{P}^{}_{e \mu}$ varies with the neutrino energy E at a fixed baseline $L = 1000$ km. We can find that the UV effects could modify the oscillation probabilities up to a few percentages while the oscillatory behaviors are not largely changed. These features can be understood with the help of Eqs.~(14) and (B6), where for the long-baseline accelerator neutrino oscillation experiments the fast oscillating terms of $\Delta^{}_{41}$ are averaged out, and both the DUV and the IUV effects could lead to corrections to the magnitudes of the oscillatory terms and small shifts of the oscillatory frequencies (if the matter effects are taken into account) which are all of the order of $s^2$.

At the baseline $L = 1000$ km, different oscillatory behaviors of the neutrino and antineutrino channels in either the NH or the IH cases are illustrated in Fig.~3. The distinct behaviors between the two kinds of mass hierarchies can be observed in the vicinity of the first oscillation maxima (at $E \sim 2$ GeV), and the differences between the neutrino and the antineutrino channels are obvious at the second peaks  (at $E \sim 0.5$ GeV). This conclusion is true no matter if the unitarity is violated and what kind of UV it is.
We study in Figs.~4 and 5 the dependence of the UV corrections to the oscillation probabilities on those UV parameters. Generally speaking, the larger the active-sterile mixing is, the more significant the UV effects would be. However, whether the oscillatory behaviors are slightly enhanced or depressed and the magnitude of the UV corrections depend on the actual active-sterile mixing patterns (the mixing angles as well as the phases).

The equations $\tilde{P}(\nu^{}_{\alpha} \rightarrow \nu^{}_{e, \mu, \tau}) = 1$ (for $\alpha = e, \mu, \tau$) always hold if there exists only three standard neutrinos and the $3 \times 3$ MNSP matrix is unitary. In the presence of UV, these summed probabilities may deviate from unity up to the order of $s^2$. The larger the UV is, the more significant the deviation would be (see Fig.~8). Moreover, this deviation may be slightly enhanced at the region where the matter effects are important. We show in Fig.~6 the contours of $\tilde{P}(\nu^{}_{\mu} \rightarrow \nu^{}_{e, \mu, \tau})$ as functions of the neutrino energy $E$ and baseline length $L$ in the (3+0+0) (top-left), (3+0+2) (bottom-left), (3+1+0) (top-right) and (3+1+2) (bottom-right) scenarios respectively. In the DUV case (in the presence of light sterile neutrinos), we always have $\tilde{P}(\nu^{}_{\mu} \rightarrow \nu^{}_{e, \mu, \tau}) \leq 1$. However in the IUV case (in the presence of only heavy sterile neutrinos) $\tilde{P}(\nu^{}_{\mu} \rightarrow \nu^{}_{e, \mu, \tau})$ can be either smaller than 1 or slightly larger than 1. This can be easily understood by using Eqs.~(17) - (19), where the suppression terms $(1-2s^2_{i4})$ dominate in the case of DUV.

To see the intrinsic properties of $\tilde{P}(\nu^{}_{\mu} \rightarrow \nu^{}_{e, \mu, \tau})$ more clearly, we illustrate in Figs.~7 - 9 the one-dimensional plots of this probability as functions of neutrino energy $E$ at the fixed baseline of $L=1000$ km. Different mass hierarchy scenarios are compared in Fig.~7. One can find the curve of the NH case and that of the IH case are approximately horizontally-symmetric, since the hierarchy-asymmetric terms dominate the energy dependency as shown in Eqs.~(17) - (19). Suppose three masses of the active neutrinos are of the normal hierarchy, Fig.~9 shows the probabilities $\tilde{P}(\nu^{}_{\mu} \rightarrow \nu^{}_{e, \mu, \tau})$ with different choices of the UV CP-violating phases. In comparison with Fig.~7, we can easily find that if the sign of $\Delta m^{2}_{31}$ and the signs of all $\delta^{}_{ij}$ (for $i = 1, 2, 3$ and $j = 4, 5, 6$) flipped at the same time, the probabilities $\tilde{P}(\nu^{}_{\mu} \rightarrow \nu^{}_{e, \mu, \tau})$ (for $\alpha = e, \mu, \tau$) would approximately have the same values. This results can also be easily inferred from Eqs.~(26) - (28).

Before finishing this section, we shall test the accuracy of our approximate analytical formulas in Eqs.~(11) - (16). We show in Fig.~10 the absolute errors (defined as the absolute differences between the analytical and the numerical calculations of the oscillation probabilities, i.e., $\left | \tilde{P}^{\rm analytical}_{\alpha \beta} - \tilde{P}^{\rm numerical}_{\alpha \beta} \right |$) as functions of the neutrino energy $E$ and the baseline length $L$. We can find from the figure that, in the region $L / E \ll 10^3$ km/GeV, these analytical formulas are of good accuracy (with the absolute errors better than 0.01). Therefore, we conclude that the series expansions of the oscillation probabilities where the UV effects have been taken into account are useful for understanding of the physics of future long-baseline neutrino oscillation experiments, which might be sensitive to the UV effects.

\section{Conclusion}
The purpose of this paper is to derive the approximate analytical formulas for neutrino oscillation probabilities in matter of constant density in the presence of the direct and the indirect unitarity violation. The probabilities are expanded in both the mass hierarchy parameter $\alpha \equiv \Delta m^{2}_{21} / \Delta m^{2}_{31}$ and the small active-sterile mixing parameters $s^{2}_{ij}$ (for $i = 1, 2, 3$ and $j = 4, 5, 6$). We have showed that both the charged-current and the neutral-current interactions are non-trivial in the presence of UV. Equations (11) - (19) constitute the main analytical conclusions of this paper. We have discussed the important properties of the oscillation probabilities by comparing among scenarios of different mass hierarchies, different mixing patterns and different CP-violating phases. The accuracy of these analytical expressions in different regions of the neutrino energy $E$ and the baseline length $L$ are also studied to identify the valid region for realistic applications.
Our numerical calculations show that, for most of practical long-baseline accelerator experiments, these analytical formulas are very good approximations.
In particular, we have discussed the deviations of the ``to all'' probabilities $\tilde{P}(\nu^{}_{\alpha} \rightarrow \nu^{}_{e, \mu, \tau})$ from the unity, which provides a definite signal of the UV. We find that the behaviors of these summed probabilities are quite different in the case of DUV or IUV. However there is an approximate degeneracy if the signs of $\Delta m^{2}_{31}$ and all the UV CP-violating phases flipped at the same time. In the presence of UV, the deviations can be enhanced by the matter effects, where only the charged-current interaction is relevant.

As the measurements of neutrino oscillations enter the precision era, testing the standard three-neutrino paradigm becomes a potential probe for new physics beyond the standard neutrino oscillation framework. It is desirable to present a sophisticated study for the test of the UV in future long-baseline accelerator neutrino experiments or the Neutrino Factory. The analytical expressions given in this paper could reveal the basic dependence of the neutrino oscillation probabilities on the DUV or the IUV parameters and help to facilitate the choice of the experimental setup as well as the analysis of the data.

\begin{acknowledgements}
This work is supported by the National Natural Science Foundation of China under Grant Nos. 11135009, 11305193, and 11105113 , and by the CAS Center for Excellence in Particle Physics (CCEPP).

\end{acknowledgements}

\begin{appendix}

\section{Diagonalization of the Hamiltonian in matter using the perturbation theory}

In this appendix, we use the perturbation theory to diagonalize the $4 \times 4$ effective Hamiltonian $\tilde{{\cal H}}$ in matter of constant density by the unitary transformation $\tilde{{\cal H}} = X^{}_{} \tilde{E} X^{\dagger}_{}$, where $X$ is a $4 \times 4$ unitary matrix and $\tilde{E}$ is diagonal. In the series expansion of $\tilde{\cal{H}}$, we regard $\alpha \equiv \Delta m_{21}^{2} / \Delta m_{31}^{2}$ and $s^{2}_{ij}$ (for $i = 1, 2, 3$ and $j = 4, 5, 6$) as the small parameters of the same order $s^2$, and perform the diagonalization to the first order of $\alpha$ and $s^{2}_{ij}$. In the appendices, we shall adopt the following abbreviations:
\begin{eqnarray}
\Delta^{}_{31} \; \equiv \; \frac{\Delta m^{2}_{31} L}{4 E} \; , ~~~~~~
A^{}_{CC} \; \equiv \; \frac{2 E^{}_{\nu} V^{}_{CC}}{\Delta m^{2}_{31}} \; , ~~~~~~
A^{}_{NC} \; \equiv \; \frac{2 E^{}_{\nu} V^{}_{NC}}{\Delta m^{2}_{31}} \; ,
\end{eqnarray}
\begin{eqnarray}
T^{}_{11} & \equiv & s^2_{14} + s^2_{15} + s^2_{16} \; , \nonumber\\
T^{}_{22} & \equiv & s^2_{24} + s^2_{25} + s^2_{26} \; , \nonumber\\
T^{}_{33} & \equiv & s^2_{34} + s^2_{35} + s^2_{36} \; , \nonumber\\
R^{}_{12} & \equiv & {\rm Re} \left [ \left ( \hat{s}^{}_{14} \hat{s}^{*}_{24} + \hat{s}^{}_{15} \hat{s}^{*}_{25} + \hat{s}^{}_{16} \hat{s}^{*}_{26} \right ) e^{-i\delta} \right ] \; , \nonumber\\
R^{}_{13} & \equiv & {\rm Re} \left [ \left ( \hat{s}^{}_{14} \hat{s}^{*}_{34} + \hat{s}^{}_{15} \hat{s}^{*}_{35} + \hat{s}^{}_{16} \hat{s}^{*}_{36} \right ) e^{-i\delta} \right ] \; , \nonumber\\
R^{}_{23} & \equiv & {\rm Re} \left [ \hat{s}^{}_{24} \hat{s}^{*}_{34} + \hat{s}^{}_{25} \hat{s}^{*}_{35} + \hat{s}^{}_{26} \hat{s}^{*}_{36} \right ] \; , \nonumber\\
I^{}_{12} & \equiv & {\rm Im} \left [ \left ( \hat{s}^{}_{14} \hat{s}^{*}_{24} + \hat{s}^{}_{15} \hat{s}^{*}_{25} + \hat{s}^{}_{16} \hat{s}^{*}_{26} \right ) e^{-i\delta} \right ] \; , \nonumber\\
I^{}_{13} & \equiv & {\rm Im} \left [ \left ( \hat{s}^{}_{14} \hat{s}^{*}_{34} + \hat{s}^{}_{15} \hat{s}^{*}_{35} + \hat{s}^{}_{16} \hat{s}^{*}_{36} \right ) e^{-i\delta} \right ] \; , \nonumber\\
I^{}_{23} & \equiv & {\rm Im} \left [ \hat{s}^{}_{24} \hat{s}^{*}_{34} + \hat{s}^{}_{25} \hat{s}^{*}_{35} + \hat{s}^{}_{26} \hat{s}^{*}_{36} \right ] \; ,
\end{eqnarray}
and
\begin{eqnarray}
T^{}_{a} & \equiv & R^{}_{23} \cos2\theta^{}_{23} + \frac{1}{2} \left ( T^{}_{22} - T^{}_{33} \right ) \sin2\theta^{}_{23} \; , \nonumber\\
T^{}_{b} & \equiv & R^{}_{23} \sin2\theta^{}_{23} + T^{}_{22} s^{2}_{23} + T^{}_{33} c^{2}_{23} \; , \nonumber\\
T^{}_{c} & \equiv & 2 R^{}_{23} \sin2\theta^{}_{23} + \left ( T^{}_{22} - T^{}_{33} \right ) \cos2\theta^{}_{23} \; , \nonumber\\
R^{}_{a} & \equiv & - R^{}_{12} c^{}_{23} + R^{}_{13} s^{}_{23} \; , \nonumber\\
R^{}_{b} & \equiv & R^{}_{12} s^{}_{23} + R^{}_{13} c^{}_{23} \; , \nonumber\\
R^{}_{c} & \equiv & R^{}_{12} c^{}_{23} + R^{}_{13} s^{}_{23} \; , \nonumber\\
I^{}_{a} & \equiv & - I^{}_{12} c^{}_{23} + I^{}_{13} s^{}_{23} \; , \nonumber\\
I^{}_{b} & \equiv & I^{}_{12} s^{}_{23} + I^{}_{13} c^{}_{23} \; .
\end{eqnarray}
In the basis of neutrino mass eigenstates in vacuum, $\tilde{\cal{H}}$ can be written as
\begin{eqnarray}
\tilde{\cal{H}} \; = \; E + U^{T}_{} \bar{A} U^{*}_{} \; \simeq \; E_{1}^{} \cdot \mathbf{1} + \frac{1}{2E_{\nu}^{}} \cdot {\rm diag} \left ( 0, \; \alpha, \; \Delta m^{2}_{31}, \; \Delta m^{2}_{41} \right ) + U^{T}_{} \bar{A} U^{*}_{} \; ,
\end{eqnarray}
where $\bar{A} \equiv {\rm diag} \left \{ V^{}_{CC} - V^{}_{NC}, \; - V^{}_{NC}, \; - V^{}_{NC} \right \}$, and $U$ is just the $3 \times 4$ non-unitary mixing matrix in vacuum. The matrix $\tilde{U}$, that describes the effective neutrino mixing in matter, can be expressed as $\tilde{U} = U X^{*}_{}$ which is also a non-unitary $3 \times 4$ matrix.

We find that it is easier to perform the diagonalization of $\tilde{{\cal H'}} \equiv U^{*}_{0}
\tilde{{\cal H}} U^{T}_{0}$, with
\begin{eqnarray}
U^{}_{0} \; = \; \left ( \begin{matrix} ~ V^{}_{0} ~ & \cr & ~ 1 ~ \end{matrix} \right ) \; ,
\end{eqnarray}
by doing the unitary transformation $\tilde{{\cal H}}' = X'^{}_{} \tilde{E} X'^{\dagger}_{}$, and therefore the effective neutrino mixing matrix in matter is
\begin{eqnarray}
\tilde{U} & = & U \left ( U^{T}_{0} X' \right )^{*}_{} \; = \; U U^{\dagger}_{0} X'^{*}_{} \nonumber\\
& \approx & \left ( \begin{matrix} 1 - \displaystyle\frac{1}{2} \left ( s^{2}_{14} + s^{2}_{15} + s^{2}_{16} \right ) & 0 & 0 & \hat{s}^{*}_{14} \cr - \hat{s}^{}_{14} \hat{s}^{*}_{24} + \hat{s}^{}_{15} \hat{s}^{*}_{25} + \hat{s}^{}_{16} \hat{s}^{*}_{26} & 1 - \displaystyle\frac{1}{2} \left ( s^{2}_{24} + s^{2}_{25} + s^{2}_{26} \right ) & 0 & \hat{s}^{*}_{24} \cr - \hat{s}^{}_{14} \hat{s}^{*}_{34} + \hat{s}^{}_{15} \hat{s}^{*}_{35} + \hat{s}^{}_{16} \hat{s}^{*}_{36} & - \hat{s}^{}_{24} \hat{s}^{*}_{34} + \hat{s}^{}_{25} \hat{s}^{*}_{35} + \hat{s}^{}_{26} \hat{s}^{*}_{36} & 1 - \displaystyle\frac{1}{2} \left ( s^{2}_{34} + s^{2}_{35} + s^{2}_{36} \right ) & \hat{s}^{*}_{34} \end{matrix} \right ) X'^{*}_{} \; . \nonumber\\
\end{eqnarray}
In this new basis, we can write down the series expansion of the Hamiltonian as
\begin{eqnarray}
\tilde{\cal{H'}} \; = \; \tilde{\cal{H'}}^{(0)}_{} + \tilde{\cal{H'}}^{(1)}_{} + ... \; ,
\end{eqnarray}
where $\tilde{\cal{H'}}^{(0)}_{}$ and $\tilde{\cal{H'}}^{(1)}_{}$ can be found in Eq.~(A8) and Eq~(A9), respectively.
For the eigenvalues and eigenvectors, we also write as $\tilde{E'}_{i}^{} = \tilde{E}_{i}^{(0)} + \tilde{E}_{i}^{(1)} + ... $ and $v_{i}^{} = v_{i}^{(0)} + v_{i}^{(1)} + ... $ (for $i = 1,2,3,4$), and the unitary matrix is defined as $X' = \left ( v_{1}^{}, v_{2}^{}, v_{3}^{}, v_{4}^{} \right )$.

\newpage
\begin{landscape}
\thispagestyle{empty}
\begin{eqnarray}
\tilde{\cal{H'}}^{(0)}_{} & = & E^{}_{1} \cdot \mathbf{1} + \frac{\Delta m^{2}_{31}}{2E_{\nu}^{}} \left ( \begin{matrix} s^{2}_{13} & s^{}_{23} s^{}_{13} c^{}_{13} e^{i\delta}_{} & c^{}_{23} s^{}_{13} c^{}_{13} e^{i\delta}_{} & 0 \cr s^{}_{23} s^{}_{13} c^{}_{13} e^{-i\delta}_{} & s^{2}_{23} c^{2}_{13} & s^{}_{23} c^{}_{23} c^{2}_{13} & 0 \cr c^{}_{23} s^{}_{13} c^{}_{13} e^{-i\delta}_{} & s^{}_{23} c^{}_{23} c^{2}_{13} & c^{2}_{23} c^{2}_{13} & 0 \cr 0 & 0 & 0 & \Delta m^{2}_{41} / \Delta m^{2}_{31} \end{matrix} \right ) + \left ( \begin{matrix} V^{}_{CC} - V^{}_{NC} & & & \cr & - V^{}_{NC} & & \cr & & - V^{}_{NC} & \cr & & & ~ 0 ~ \end{matrix} \right ) \; , \\
\nonumber\\
\tilde{\cal{H'}}^{(1)}_{} & = & \alpha \frac{\Delta m^{2}_{31}}{2E_{\nu}^{}} \left ( \begin{matrix} s^{2}_{12} c^{2}_{13} & \left ( c^{}_{12} c^{}_{23} - s^{}_{12} s^{}_{23} s^{}_{13} e^{i\delta}_{} \right ) s^{}_{12} c^{}_{13} & - \left ( c^{}_{12} s^{}_{23} + s^{}_{12} c^{}_{23} s^{}_{13} e^{i\delta}_{} \right ) s^{}_{12} c^{}_{13} & 0 \cr \left ( c^{}_{12} c^{}_{23} - s^{}_{12} s^{}_{23} s^{}_{13} e^{-i\delta}_{} \right ) s^{}_{12} c^{}_{13} & \begin{matrix} c^{2}_{12} c^{2}_{23} + s^{2}_{12} s^{2}_{23} s^{2}_{13} \cr - \frac{1}{2} \sin2\theta^{}_{12} \sin2\theta^{}_{23} s^{}_{13} \cos\delta \end{matrix} & \begin{matrix} \frac{1}{2} \sin2\theta^{}_{23} \left( s^{2}_{12} s^{2}_{13} - c^{2}_{12} \right ) \cr - \frac{1}{2} \sin2\theta^{}_{12} s^{}_{13} \left ( \cos2\theta^{}_{23} \cos\delta + i \sin\delta \right ) \end{matrix} & 0 \cr - \left ( c^{}_{12} s^{}_{23} + s^{}_{12} c^{}_{23} s^{}_{13} e^{-i\delta}_{} \right ) s^{}_{12} c^{}_{13} &  \begin{matrix} \frac{1}{2} \sin2\theta^{}_{23} \left( s^{2}_{12} s^{2}_{13} - c^{2}_{12} \right ) \cr - \frac{1}{2} \sin2\theta^{}_{12} s^{}_{13} \left ( \cos2\theta^{}_{23} \cos\delta - i \sin\delta \right ) \end{matrix} & \begin{matrix} c^{2}_{12} s^{2}_{23} + s^{2}_{12} c^{2}_{23} s^{2}_{13} \cr + \frac{1}{2} \sin2\theta^{}_{12} \sin2\theta^{}_{23} s^{}_{13} \cos\delta \end{matrix} & 0 \cr 0 & 0 & 0 & ~ 0 ~ \end{matrix} \right ) \nonumber\\
& & + V^{}_{NC} \left ( \begin{matrix} ~ s^{2}_{14} + s^{2}_{15} + s^{2}_{16} ~ & ~ \hat{s}^{}_{14} \hat{s}^{*}_{24} + \hat{s}^{}_{15} \hat{s}^{*}_{25} + \hat{s}^{}_{16} \hat{s}^{*}_{26} ~ & ~ \hat{s}^{}_{14} \hat{s}^{*}_{34} + \hat{s}^{}_{15} \hat{s}^{*}_{35} + \hat{s}^{}_{16} \hat{s}^{*}_{36} ~ & ~ - \hat{s}^{}_{14} ~ \cr ~ \hat{s}^{*}_{14} \hat{s}^{}_{24} + \hat{s}^{*}_{15} \hat{s}^{}_{25} + \hat{s}^{*}_{16} \hat{s}^{}_{26} ~  & ~ s^{2}_{24} + s^{2}_{25} + s^{2}_{26} ~ & ~ \hat{s}^{}_{24} \hat{s}^{*}_{34} + \hat{s}^{}_{25} \hat{s}^{*}_{35} + \hat{s}^{}_{26} \hat{s}^{*}_{36} ~ & ~ - \hat{s}^{}_{24} ~ \cr ~ \hat{s}^{*}_{14} \hat{s}^{}_{34} + \hat{s}^{*}_{15} \hat{s}^{}_{35} + \hat{s}^{*}_{16} \hat{s}^{}_{36} ~ & ~ \hat{s}^{*}_{24} \hat{s}^{}_{34} + \hat{s}^{*}_{25} \hat{s}^{}_{35} + \hat{s}^{*}_{26} \hat{s}^{}_{36} ~ & ~ s^{2}_{34} + s^{2}_{35} + s^{2}_{36} ~ & ~ - \hat{s}^{}_{34} ~ \cr ~ - \hat{s}^{*}_{14} ~ & ~ - \hat{s}^{*}_{24} ~ & ~ - \hat{s}^{*}_{34} ~ & ~ - \left ( s^{2}_{14} + s^{2}_{24} + s^{2}_{34} \right ) ~ \end{matrix} \right ) \nonumber\\
& & + V^{}_{CC} \left ( \begin{matrix} - \left ( s^{2}_{14} + s^{2}_{15} + s^{2}_{16} \right ) & ~ 0 ~ & ~ 0 ~ & ~ \hat{s}^{}_{14} \cr ~ 0 ~ & ~ 0 ~ & ~ 0 ~ & ~ 0 ~ \cr ~ 0 ~ & ~ 0 ~ & ~ 0 ~ & ~ 0 ~ \cr ~ \hat{s}^{*}_{14} & ~ 0 ~ & ~ 0 ~ & ~ s^{2}_{14} ~ \end{matrix} \right ) \; .
\end{eqnarray}
\newpage
\end{landscape}

$\tilde{\cal{H'}}_{}^{(0)}$ can be easily diagonalized. The resulting eigenvalues and the eigenvectors of $\tilde{\cal{H'}}_{}^{(0)}$ are
\begin{eqnarray}
\tilde{E}_{1}^{(0)} & = & E^{}_{1} - V^{}_{NC} \; , \nonumber\\
\tilde{E}_{2}^{(0)} & = & E^{}_{1} - V^{}_{NC} + \frac{1}{2} V^{}_{CC} + \frac{1}{2} \frac{\Delta m^{2}_{31}}{2E_{\nu}^{}} \left ( 1 - C \right ) \; , \nonumber\\
\tilde{E}_{3}^{(0)} & = & E^{}_{1} - V^{}_{NC} + \frac{1}{2} V^{}_{CC} + \frac{1}{2} \frac{\Delta m^{2}_{31}}{2E_{\nu}^{}} \left ( 1 + C \right ) \; , \nonumber\\
\tilde{E}_{4}^{(0)} & = & E^{}_{1} + \frac{\Delta m^{2}_{41}}{2E_{\nu}^{}} \; ,
\end{eqnarray}
where $C \equiv \sqrt{1 - 2 A^{}_{CC} \cos2\theta^{}_{13}}$,
and
\begin{eqnarray}
X'^{(0)}_{} & = & \left ( \begin{matrix} 0 & \tilde{c}^{}_{13} & \tilde{s}^{}_{13} e^{i\delta}_{} & 0 \cr - c^{}_{23} & - s^{}_{23} \tilde{s}^{}_{13} e^{-i\delta}_{} & s^{}_{23} \tilde{s}^{}_{13} & 0 \cr s^{}_{23} & - c^{}_{23} \tilde{s}^{}_{13} e^{-i\delta}_{} & c^{}_{23} \tilde{s}^{}_{13} & 0 \cr 0 & 0 & 0 & ~ 1 ~ \end{matrix} \right ) \; ,
\end{eqnarray}
where
\begin{eqnarray}
\tilde{s}^{}_{13} & \equiv & \sin\hat{\theta}^{}_{13} \; = \; \sqrt{\frac{1}{2} \left ( 1 - \frac{\cos2\theta^{}_{13} - A^{}_{CC}}{C} \right )} \; , \\
\tilde{c}^{}_{13} & \equiv & \cos\hat{\theta}^{}_{13} \; = \; \sqrt{\frac{1}{2} \left ( 1 + \frac{\cos2\theta^{}_{13} - A^{}_{CC}}{C} \right )} \; .
\end{eqnarray}
Furthermore, the first order corrections to the eigenvalues and eigenvectors are given by
\begin{eqnarray}
\tilde{E}^{(1)}_{i} & = & \tilde{\cal{H'}}^{(1)}_{ii} \; , \\
v^{(1)}_{i} & = & \sum^{}_{j \neq i} \frac{\tilde{\cal{H'}}^{(1)}_{ji}}{\tilde{E'}^{(0)}_{i} - \tilde{E'}^{(0)}_{j}} \cdot v_{j}^{(0)} \; ,
\end{eqnarray}
where $\tilde{\cal{H'}}^{(n)}_{ij} \equiv {v^{(0)}_{i}}^{\dagger} \tilde{\cal{H'}}^{(n)}_{} v^{(0)}_{j}$.
Inserting Eqs.~(A9) - (A11) into Eqs.~(A14) and (A15), we can obtain the analytical expressions of $\tilde{E}_{i}^{(1)}$
\begin{eqnarray}
\tilde{E}_{1}^{(1)} & = & \alpha \frac{\Delta m^{2}_{31}}{2E_{\nu}^{}} c^{2}_{12} + V^{}_{NC} \left ( T^{}_{22} c^{2}_{23} + T^{}_{33} s^{2}_{23} - R^{}_{23} \sin2\theta^{}_{23} \right )  \; , \nonumber\\
\tilde{E}_{2}^{(1)} & = & \frac{1}{2} \alpha \frac{\Delta m^{2}_{31}}{2E_{\nu}^{}} s^{2}_{12} \left ( 1 + \frac{1 - A^{}_{CC} \cos2\theta^{}_{13}}{C} \right ) + \left ( V^{}_{NC} - V^{}_{CC} \right ) T^{}_{11} \tilde{c}^{2}_{13} + V^{}_{NC} \left ( T^{}_{b} \tilde{s}^{2}_{13} - R^{}_{b} \sin2\tilde{\theta}^{}_{13} \right ) \; , \nonumber\\
\tilde{E}_{3}^{(1)} & = & \frac{1}{2} \alpha \frac{\Delta m^{2}_{31}}{2E_{\nu}^{}} s^{2}_{12} \left ( 1 - \frac{1 - A^{}_{CC} \cos2\theta^{}_{13}}{C} \right ) + \left ( V^{}_{NC} - V^{}_{CC} \right ) T^{}_{11} \tilde{s}^{2}_{13} + V^{}_{NC} \left ( T^{}_{b} \tilde{c}^{2}_{13} + R^{}_{b} \sin2\tilde{\theta}^{}_{13} \right ) \; , \nonumber\\
\tilde{E}_{4}^{(1)} & = & - \left ( V^{}_{NC} - V^{}_{CC} \right ) s^{2}_{14} - V^{}_{NC} \left ( s^{2}_{24} + s^{2}_{34} \right ) \ \; ,
\end{eqnarray}
as well as the expression of $X'^{(1)}_{}$
\begin{eqnarray}
X'^{(1)}_{} & = & \alpha \sin2\theta^{}_{12} \left ( \begin{matrix} \displaystyle \frac{1}{2 A^{}_{CC} c^{}_{13}} & 0 & 0 & 0 \cr \displaystyle - \frac{ \left( 1 + A^{}_{CC} \right ) s^{}_{23} s^{}_{13} e^{-i\delta}_{}}{2 A^{}_{CC} c^{2}_{13}} & \displaystyle \frac{\left ( \tilde{s}^{}_{13} s^{}_{13} + \tilde{c}^{}_{13} c^{}_{13} \right ) c^{}_{23}}{A^{}_{CC} + \left ( 1 - C \right )} & \displaystyle - \frac{\left ( \tilde{c}^{}_{13} s^{}_{13} - \tilde{s}^{}_{13} c^{}_{13} \right ) c^{}_{23} e^{i\delta}_{}}{A^{}_{CC} + \left ( 1 + C \right )} & 0 \cr \displaystyle - \frac{ \left( 1 + A^{}_{CC} \right ) c^{}_{23} s^{}_{13} e^{-i\delta}_{}}{2 A^{}_{CC} c^{2}_{13}} & \displaystyle - \frac{\left ( \tilde{s}^{}_{13} s^{}_{13} + \tilde{c}^{}_{13} c^{}_{13} \right ) s^{}_{23}}{A^{}_{CC} + \left ( 1 - C \right )} & \displaystyle \frac{\left ( \tilde{c}^{}_{13} s^{}_{13} - \tilde{s}^{}_{13} c^{}_{13} \right ) s^{}_{23} e^{i\delta}_{}}{A^{}_{CC} + \left ( 1 + C \right )} & 0 \cr 0 & 0 & 0 &0\end{matrix} \right ) \nonumber\\
& & - \frac{\alpha A^{}_{CC} s^{2}_{12} \sin2\theta^{}_{13} e^{-i\delta}_{}}{2 C^{2}_{}} \left ( \begin{matrix} ~ 0 ~ & \tilde{s}^{}_{13} & -\tilde{c}^{}_{13} e^{i\delta}_{} & 0 \cr 0 & s^{}_{23} \tilde{c}^{}_{13} e^{-i\delta}_{} & s^{}_{23} \tilde{s}^{}_{13} & 0 \cr 0 & c^{}_{23} \tilde{c}^{}_{13} e^{-i\delta}_{} & c^{}_{23} \tilde{s}^{}_{13} & 0 \cr 0 & 0 & 0 & ~ 0 ~ \end{matrix} \right ) \; .
\end{eqnarray}
Since the light sterile neutrino mass is supposed to be around the eV scale, we have safely omitted the terms proportional to $s^{}_{} \Delta m^{2}_{31} / \Delta m^{2}_{41}$. The effective neutrino mass-squared differences $\Delta \tilde{m}^{2}_{21}$, $\Delta \tilde{m}^{2}_{31}$, $\Delta \tilde{m}^{2}_{32}$ can then be written as:
\begin{eqnarray}
\Delta \tilde{m}^{2}_{21} & = & \Delta m^{2}_{31} \left \{ \frac{A^{}_{CC} + \left ( 1 - C \right )}{2} - \alpha \left [ c^{2}_{12} + s^{2}_{12} \frac{A^{}_{CC} \cos2\theta^{}_{13} - \left ( 1 + C \right )}{2C} \right ] + \left ( A^{}_{NC} - A^{}_{CC} \right ) T^{}_{11} \tilde{c}^{2}_{13}  \right. \nonumber\\
& & \left. - A^{}_{NC} \left [ T^{}_{22} \left ( c^{2}_{23} - s^{2}_{23} \tilde{s}^{2}_{13} \right ) + T^{}_{33} \left ( s^{2}_{23} - c^{2}_{23} \tilde{s}^{2}_{13} \right ) - R^{}_{23} \sin2\theta^{}_{23} \left ( 1 + \tilde{s}^{2}_{13} \right ) + R^{}_{b} \sin2\tilde{\theta}^{}_{13} \right ]\right \} \; , \nonumber\\
\Delta \tilde{m}^{2}_{31} & = & \Delta m^{2}_{31} \left \{ \frac{A^{}_{CC} + \left ( 1 + C \right )}{2} - \alpha \left [ c^{2}_{12} - s^{2}_{12} \frac{A^{}_{CC} \cos2\theta^{}_{13} - \left ( 1 - C \right )}{2C} \right ] + \left ( A^{}_{NC} - A^{}_{CC} \right ) T^{}_{11} \tilde{s}^{2}_{13}  \right. \nonumber\\
& & \left. - A^{}_{NC} \left [ T^{}_{22} \left ( s^{2}_{23} - c^{2}_{23} \tilde{s}^{2}_{13} \right ) + T^{}_{33} \left ( c^{2}_{23} - s^{2}_{23} \tilde{s}^{2}_{13} \right ) - R^{}_{23} \sin2\theta^{}_{23} \left ( 1 + \tilde{c}^{2}_{13} \right ) - R^{}_{b} \sin2\tilde{\theta}^{}_{13} \right ]\right \} \; , \nonumber\\
\Delta \tilde{m}^{2}_{32} & = & \Delta m^{2}_{31} \left [ C - \alpha s^{2}_{12} \frac{1 - A^{}_{CC} \cos2\theta^{}_{13}}{C} - \left ( A^{}_{NC} - A^{}_{CC} \right ) T^{}_{11} \cos2\tilde{\theta}^{}_{13} \right. \nonumber\\
& & \left. + A^{}_{NC} \left ( T^{}_{b} \cos2\tilde{\theta}^{}_{13} + 2 R^{}_{b} \sin2\tilde{\theta}^{}_{13} \right ) \right ] \; ,
\end{eqnarray}
and the effective mixing matrix $\tilde{U}$ in matter can then be calculated by Eq.~(A6).
Note that in this paper we do not order the eigenvalues of $\tilde{\cal H}$ according to their magnitude and the mass spectrum, since the ordering does not change the neutrino oscillation probabilities.

\section{Neutrino oscillation probabilities in matter}

In terms of effective neutrino masses and mixing in matter, the neutrino oscillation probabilities in matter $\tilde{P} ( \stackrel{(-)}{\nu}^{}_{\alpha} \rightarrow \stackrel{(-)}{\nu}^{}_{\beta} )$ have the identical form as that in vacuum:
\begin{eqnarray}
\tilde{P} ( \stackrel{(-)}{\nu}^{}_{\alpha} \rightarrow \stackrel{(-)}{\nu}^{}_{\beta} ) & = & \frac{1}{\left ( \sum^{}_{i=1,2,3,4} | \tilde{U}^{}_{\alpha i} |^2 \right ) \left ( \sum^{}_{i=1,2,3,4} | \tilde{U}^{}_{\beta i} |^2 \right )} \left  \{ \left | \sum^{}_{i=1,2,3,4} \tilde{U}^{*}_{\alpha i} \tilde{U}^{}_{\beta i} \right |^2 \right. \nonumber\\
&-& \left. 4 \sum^{}_{j > i} {\rm Re} \left [ \tilde{U}^{}_{\alpha i} \tilde{U}^{}_{\beta j} \tilde{U}^{*}_{\alpha j} \tilde{U}^{*}_{\beta i} \right ] \sin^2 \tilde{\Delta}^{}_{ji} \pm 2 \sum^{}_{j > i} {\rm Im} \left [ \tilde{U}^{}_{\alpha i} \tilde{U}^{}_{\beta j} \tilde{U}^{*}_{\alpha j} \tilde{U}^{*}_{\beta i} \right ] \sin 2 \tilde{\Delta}^{}_{ji} \right \}
\end{eqnarray}
where $\tilde{\Delta}^{}_{ji} = \Delta \tilde{m}^{2}_{ji} L / 4 E^{}_{\nu}$ with $\Delta \tilde{m}^2_{ji} \equiv \tilde{m}^{2}_{j} - \tilde{m}^{2}_{i} = 2 E^{}_{\nu} ( \tilde{E}^{}_{j} - \tilde{E}^{}_{i} )$, and the lower sign is for the antineutrino oscillation probabilities. It should be noted that the mass indices $i$, $j$ run over only the light neutrinos which can be kinematically produced in low energy neutrino oscillation experiments. For the $\Delta m^{2}_{31}$-dominated accelerator neutrino oscillation experiments, oscillatory behaviors of the terms driven by $\Delta m^{2}_{4i}$ (for $i = 1,2,3$) are averaged out. In this case the oscillation probabilities can be approximately written as
\begin{eqnarray}
\tilde{P} \; ( \stackrel{(-)}{\nu}^{}_{\alpha} \rightarrow  \stackrel{(-)}{\nu}^{}_{\beta} ) & \approx & \frac{1}{\left ( \sum^{}_{i=1,2,3,4} | \tilde{U}^{}_{\alpha i} |^2 \right ) \left ( \sum^{}_{i=1,2,3,4} | \tilde{U}^{}_{\beta i} |^2 \right )} \left  \{ \left | \sum^{}_{i=1,2,3,4} \tilde{U}^{*}_{\alpha i} \tilde{U}^{}_{\beta i} \right |^2 \right. \nonumber\\
& & - 4 {\rm Re} \left [ \tilde{U}^{}_{\alpha 1} \tilde{U}^{}_{\beta 2} \tilde{U}^{*}_{\alpha 2} \tilde{U}^{*}_{\beta 1} \right ] \sin^2 \tilde{\Delta}^{}_{21} \pm 2 {\rm Im} \left [ \tilde{U}^{}_{\alpha 1} \tilde{U}^{}_{\beta 2} \tilde{U}^{*}_{\alpha 2} \tilde{U}^{*}_{\beta 1} \right ] \sin2\tilde{\Delta}^{}_{21} \nonumber\\
& & - 4 {\rm Re} \left [ \tilde{U}^{}_{\alpha 1} \tilde{U}^{}_{\beta 3} \tilde{U}^{*}_{\alpha 3} \tilde{U}^{*}_{\beta 1} \right ] \sin^2 \tilde{\Delta}^{}_{31} \pm 2 {\rm Im} \left [ \tilde{U}^{}_{\alpha 1} \tilde{U}^{}_{\beta 3} \tilde{U}^{*}_{\alpha 3} \tilde{U}^{*}_{\beta 1} \right ] \sin2\tilde{\Delta}^{}_{31} \nonumber\\
& & - 4 {\rm Re} \left [ \tilde{U}^{}_{\alpha 2} \tilde{U}^{}_{\beta 3} \tilde{U}^{*}_{\alpha 3} \tilde{U}^{*}_{\beta 2} \right ] \sin^2 \tilde{\Delta}^{}_{32} \pm 2 {\rm Im} \left [ \tilde{U}^{}_{\alpha 2} \tilde{U}^{}_{\beta 3} \tilde{U}^{*}_{\alpha 3} \tilde{U}^{*}_{\beta 2} \right ] \sin2\tilde{\Delta}^{}_{32} \nonumber\\
& & \left. - 2 {\rm Re} \left [ \tilde{U}^{*}_{\alpha 4} \tilde{U}^{}_{\beta 4} \left ( \tilde{U}^{}_{\alpha 1} \tilde{U}^{*}_{\beta 1} + \tilde{U}^{}_{\alpha 2} \tilde{U}^{*}_{\beta 2} + \tilde{U}^{}_{\alpha 3} \tilde{U}^{*}_{\beta 3} \right ) \right ]  \right \} \; .
\end{eqnarray}
By inserting the results obtained in Appendix A, we are able to write down the neutrino oscillation probabilities $\tilde{P}^{}_{\alpha \beta} \equiv \tilde{P} ( \nu^{}_{\alpha} \rightarrow \nu^{}_{\beta} )$ to the first order in both $\alpha \equiv \Delta m_{21}^{2} / \Delta m_{31}^{2}$ and $s^{2}_{ij}$ (for $i = 1, 2, 3$ and $j = 4, 5, 6$):
\begin{eqnarray}
\tilde{P}^{}_{e e} & \approx & 1 - \frac{1}{1 - 2 \left ( s^{2}_{15} + s^{2}_{16} \right )} \left  \{ 2 s^{2}_{14} + \left [ \sin^2 2 \tilde{\theta}^{}_{13} \left ( 1 - 2 T^{}_{11} + 2 \alpha s^{2}_{12} \cos2\tilde{\theta}^{}_{13} \frac{A^{}_{CC}}{C} \right ) \right. \right. \nonumber\\[3mm]
& & \left. + \sin4\tilde{\theta}^{}_{13} \left ( \left( 2 R^{}_{b} \cos2\tilde{\theta}^{}_{13} - T^{}_{b} \sin2\tilde{\theta}^{}_{13} \right ) \frac{A^{}_{NC}}{C} + T^{}_{11} \sin2\tilde{\theta}^{}_{13} \frac{A^{}_{NC} - A^{}_{CC}}{C} \right ) \right ] \sin^2(C \Delta^{}_{31}) \nonumber\\[3mm]
& & + \Delta^{}_{31} \sin^2 2\tilde{\theta}^{}_{13} \left [ - 2 \alpha s^{2}_{12} \frac{1 - A^{}_{CC} \cos2\theta^{}_{13}}{C} + \left ( 2 R^{}_{b} \sin2\tilde{\theta}^{}_{13} + T^{}_{b} \cos2\tilde{\theta}^{}_{13} \right ) A^{}_{NC} \right. \nonumber\\[3mm]
& & \left. \left. + T^{}_{11} \cos2\tilde{\theta}^{}_{13} \left ( A^{}_{NC} - A^{}_{CC} \right ) \right ] \sin(C \Delta^{}_{31}) \cos(C \Delta^{}_{31}) \right \} \; , \\[3mm]
\tilde{P}^{}_{\mu \mu} & \approx & 1 - \frac{1}{1 - 2 \left ( s^{2}_{25} + s^{2}_{26} \right )} \left  \{ 2 s^{2}_{24} + \sin2\theta^{}_{23} \left [ \frac{1}{2} \sin2\theta^{}_{23} \left ( 1 - 2 T^{}_{22} \right ) \right. \right. \nonumber\\[3mm]
& & \left. - \frac{\cos2\theta^{}_{23}}{A^{}_{CC} c^{2}_{13}} \left ( \alpha (1 + A^{}_{CC}) \sin2\theta^{}_{12} s^{}_{13} \cos\delta - \left ( R^{}_{a} \sin2\theta^{}_{13} + 2 T^{}_{a} \left ( A^{}_{CC} + s^{2}_{13} \right ) \right ) A^{}_{NC} \right ) \right ] \nonumber\\[3mm]
& & \cdot \left [ 1 - \cos(1+A^{}_{CC})\Delta^{}_{31} \cos(C \Delta^{}_{31}) \right ] \nonumber\\[3mm]
& & + \left [ s^{2}_{23} \left ( 1 - 2 T^{}_{22} \right ) \left ( 2 c^{2}_{23} \cos2\tilde{\theta}^{}_{13} \sin(1+A^{}_{CC})\Delta^{}_{31} + s^{2}_{23} \sin^2 2 \tilde{\theta}^{}_{13} \sin(C \Delta^{}_{31}) \right ) \right. \nonumber\\[3mm]
& & + \frac{\alpha \sin2\theta^{}_{12} \sin2\theta^{}_{23} s^{}_{13} \cos\delta}{C A^{}_{CC}} \left ( \left [ 1 +\frac{\cos2\theta^{}_{23} }{c^{2}_{13}} \left ( A^{2}_{CC} + 2 s^{2}_{13} A^{}_{CC} + s^{2}_{13}  \right ) \right ] \sin(1+A^{}_{CC})\Delta^{}_{31}  \right. \nonumber\\[3mm]
& & \left. - 2 s^{2}_{23} \frac{ 1 - A^{}_{CC} \cos 2 \theta^{}_{13}}{C} \sin(C \Delta^{}_{31}) \right ) \nonumber\\[3mm]
& & - 2 \alpha s^{2}_{12} s^{2}_{23} \sin^2 2\tilde{\theta}^{}_{13} \frac{A^{}_{CC}}{C} \left (c^{2}_{23} \sin(1+A^{}_{CC})\Delta^{}_{31} - 2 s^{2}_{23} \cos2\tilde{\theta}^{}_{13} \sin(C \Delta^{}_{31}) \right ) \nonumber\\[3mm]
& & - 4 R^{}_{12} s^{}_{23} \sin2\tilde{\theta}^{}_{13} \left ( c^{2}_{23} \sin(1+A^{}_{CC})\Delta^{}_{31} - s^{2}_{23} \cos2\tilde{\theta}^{}_{13} \sin(C \Delta^{}_{31}) \right ) \nonumber\\[3mm]
& & - R^{}_{a} \frac{A^{}_{NC} \sin2\theta^{}_{23} \sin2\tilde{\theta}^{}_{13}}{A^{}_{CC} c^{2}_{13}} \left ( \left [ c^{2}_{13} + \cos2\theta^{}_{23} \left ( A^{}_{CC} + s^{2}_{13} \right ) \right ] \sin(1+A^{}_{CC})\Delta^{}_{31} \right. \nonumber\\[3mm]
& & \left. - 2 s^{2}_{23} c^{2}_{13} \frac{1 - A^{}_{CC}}{C} \sin(C \Delta^{}_{31}) \right ) \nonumber\\[3mm]
& & - T^{}_{a} \frac{A^{}_{NC} \sin2\theta^{}_{23}}{C A^{}_{CC} c^{2}_{13}} \left ( \left [ \frac{1}{2} \sin^2 2\theta^{}_{13} + \cos2\theta^{}_{23} \left ( 2 s^{4}_{13} + A^{}_{CC} - 3 A^{}_{CC} \cos2\theta^{}_{13} + 2 A^{2}_{CC} \right ) \right ] \right. \nonumber\\[3mm]
& & \left. \cdot \sin(1+A^{}_{CC})\Delta^{}_{31} - s^{2}_{23} \sin^2 2\theta^{}_{13} \frac{1 + A^{}_{CC}}{C} \sin(C \Delta^{}_{31}) \right ) \nonumber\\[3mm]
& & - 4 s^{2}_{23} \sin2\tilde{\theta}^{}_{13} \left [ \left ( R^{}_{b} \cos2\tilde{\theta}^{}_{13} - \frac{1}{2} T^{}_{b} \sin2\tilde{\theta}^{}_{13} \right ) \frac{A^{}_{NC}}{C} + \frac{1}{2} T^{}_{11} \sin2\tilde{\theta}^{}_{13} \frac{A^{}_{NC} - A^{}_{CC}}{C} \right ] \nonumber\\[3mm]
& & \left. \cdot \left ( c^{2}_{23} \sin(1+A^{}_{CC})\Delta^{}_{31} - s^{2}_{23} \cos2\tilde{\theta}^{}_{13} \sin(C \Delta^{}_{31}) \right ) \right ] \sin(C \Delta^{}_{31}) \nonumber\\[3mm]
& & + \Delta^{}_{31} \sin^2 2\theta^{}_{23} \left [ \alpha \left ( s^{2}_{12} \left ( s^{2}_{13} + \frac{1}{2} \sin^2 2\tilde{\theta}^{}_{13} A^{}_{CC} \right ) - c^{2}_{12} \right ) \right. \nonumber\\[3mm]
& & + \frac{1}{2} \left ( R^{}_{b} \sin2\tilde{\theta}^{}_{13} \cos2\tilde{\theta}^{}_{13} - \frac{1}{2} T^{}_{b} \sin^2 2\tilde{\theta}^{}_{13} + T^{}_{c} \right ) A^{}_{NC} \nonumber\\[3mm]
& & \left. + \frac{1}{4} T^{}_{11} \sin^2 2\tilde{\theta}^{}_{13} \left ( A^{}_{NC} - A^{}_{CC} \right ) \right ] \sin(1+A^{}_{CC})\Delta^{}_{31} \cos(C \Delta^{}_{31}) \nonumber\\[3mm]
& & - \Delta^{}_{31} \sin^2 2\theta^{}_{23} \left [ \alpha \left ( s^{2}_{12} s^{2}_{13} \frac{1+A^{}_{CC}}{C} + c^{2}_{12} \cos2\tilde{\theta}^{}_{13} \right ) \right. \nonumber\\[3mm]
& & \left. - \frac{1}{2} \left ( R^{}_{b} \sin2\tilde{\theta}^{}_{13} + T^{}_{c} \cos2\tilde{\theta}^{}_{13} \right ) A^{}_{NC} \right ] \cos(1+A^{}_{CC})\Delta^{}_{31} \sin(C \Delta^{}_{31}) \nonumber\\[3mm]
& & - 2 \Delta^{}_{31} s^{4}_{23} \sin^2 2\tilde{\theta}^{}_{13} \left [ \alpha s^{2}_{12} \frac{1 - A^{}_{CC} \cos2\theta^{}_{13}}{C} - \left ( R^{}_{b} \sin2\tilde{\theta}^{}_{13} + \frac{1}{2} T{}_{b} \cos2\tilde{\theta}^{}_{13} \right ) A^{}_{NC} \right. \nonumber\\[3mm]
& & \left. \left. + \frac{1}{2} T^{}_{11} \cos2\tilde{\theta}^{}_{13} \left ( A^{}_{NC} - A^{}_{CC} \right ) \right ] \sin(C \Delta^{}_{31}) \cos(C \Delta^{}_{31}) \right \} \; , \\[3mm]
\tilde{P}^{}_{\tau \tau} & \approx & 1 - \frac{1}{1 - 2 \left ( s^{2}_{35} + s^{2}_{36} \right )} \left  \{ 2 s^{2}_{34} + \sin2\theta^{}_{23} \left [ \frac{1}{2} \sin2\theta^{}_{23} \left ( 1 - 2 T^{}_{33} \right  ) \right. \right. \nonumber\\[3mm]
& & - \frac{\cos2\theta^{}_{23}}{A^{}_{CC} c^{2}_{13}} \left ( \alpha (1 + A^{}_{CC}) \sin2\theta^{}_{12} s^{}_{13} \cos\delta - \left ( R^{}_{a} \sin2\theta^{}_{13} + 2 T^{}_{a} \left ( A^{}_{CC} + s^{2}_{13} \right ) \right ) A^{}_{NC} \right ) \nonumber\\[3mm]
& & \left. + 2 R^{}_{23} \cos2\theta^{}_{23} \right ] \left [ 1 - \cos(1+A^{}_{CC})\Delta^{}_{31} \cos(C \Delta^{}_{31}) \right ] \nonumber\\[3mm]
& & + \left [ c^{2}_{23} \left ( 1 - 2 T^{}_{33} \right ) \left ( 2 s^{2}_{23} \cos2\tilde{\theta}^{}_{13} \sin(1+A^{}_{CC})\Delta^{}_{31} + c^{2}_{23} \sin^2 2\tilde{\theta}^{}_{13} \sin(C \Delta^{}_{31}) \right ) \right. \nonumber\\[3mm]
& & - \frac{\alpha \sin2\theta^{}_{12} \sin2\theta^{}_{23} s^{}_{13} \cos\delta}{C A^{}_{CC}} \left ( \left [ 1 - \frac{\cos2\theta^{}_{23} }{c^{2}_{13}} \left ( A^{2}_{CC} + 2 s^{2}_{13} A^{}_{CC} + s^{2}_{13} \right ) \right ] \sin(1+A^{}_{CC})\Delta^{}_{31} \right. \nonumber\\[3mm]
& & \left. -  2 c^{2}_{23} \frac{ 1 - A^{}_{CC} \cos 2 \theta^{}_{13}}{C} \sin(C \Delta^{}_{31}) \right ) \nonumber\\[3mm]
& & - 2 \alpha s^{2}_{12} c^{2}_{23} \sin^2 2\tilde{\theta}^{}_{13} \frac{A^{}_{CC}}{C} \left ( s^{2}_{23} \sin(1+A^{}_{CC})\Delta^{}_{31} - 2 c^{2}_{23} \cos2\tilde{\theta}^{}_{13} \sin(C \Delta^{}_{31}) \right ) \nonumber\\[3mm]
& & - 4 R^{}_{13} c^{}_{23} \sin2\tilde{\theta}^{}_{13} \left ( s^{2}_{23} \sin(1+A^{}_{CC})\Delta^{}_{31} - c^{2}_{23} \cos2\tilde{\theta}^{}_{13} \sin(C \Delta^{}_{31}) \right ) \nonumber\\[3mm]
& & + 2 R^{}_{23} \sin2\theta^{}_{23} \left ( \cos 2\theta^{}_{23} \cos2\tilde{\theta}^{}_{13} \sin(1+A^{}_{CC})\Delta^{}_{31} - c^{2}_{23} \sin^2 2\tilde{\theta}^{}_{13} \sin(C \Delta^{}_{31}) \right ) \nonumber\\[3mm]
& & + R^{}_{a} \frac{A^{}_{NC} \sin2\theta^{}_{23} \sin2\tilde{\theta}^{}_{13}}{A^{}_{CC} c^{2}_{13}} \left ( \left [ c^{2}_{13} - \cos2\theta^{}_{23} \left ( A^{}_{CC} + s^{2}_{13} \right ) \right ] \sin(1+A^{}_{CC})\Delta^{}_{31} \right. \nonumber\\[3mm]
& & \left. - 2 c^{2}_{23} c^{2}_{13} \frac{1 - A^{}_{CC}}{C} \sin(C \Delta^{}_{31}) \right ) \nonumber\\[3mm]
& & + T^{}_{a} \frac{A^{}_{NC} \sin2\theta^{}_{23}}{C A^{}_{CC} c^{2}_{13}} \left ( \left [ \frac{1}{2} \sin^2 2\theta^{}_{13} - \cos2\theta^{}_{23} \left ( 2 s^{4}_{13} + A^{}_{CC} - 3 A^{}_{CC} \cos2\theta^{}_{13} + 2 A^{2}_{CC} \right ) \right ] \right. \nonumber\\[3mm]
& & \left. \cdot \sin(1+A^{}_{CC})\Delta^{}_{31} - c^{2}_{23} \sin^2 2\theta^{}_{13} \frac{1 + A^{}_{CC}}{C} \sin(C \Delta^{}_{31}) \right ) \nonumber\\[3mm]
& & - 4 c^{2}_{23} \sin2\tilde{\theta}^{}_{13} \left [ \left ( R^{}_{b} \cos2\tilde{\theta}^{}_{13} - \frac{1}{2} T^{}_{b} \sin2\tilde{\theta}^{}_{13} \right ) \frac{A^{}_{NC}}{C} + \frac{1}{2} T^{}_{11} \sin2\tilde{\theta}^{}_{13} \frac{A^{}_{NC} - A^{}_{CC}}{C} \right ] \nonumber\\[3mm]
& & \left. \cdot \left ( s^{2}_{23} \sin(1+A^{}_{CC})\Delta^{}_{31} - c^{2}_{23} \cos2\tilde{\theta}^{}_{13} \sin(C \Delta^{}_{31}) \right ) \right ] \sin(C \Delta^{}_{31}) \nonumber\\[3mm]
& & + \Delta^{}_{31} \sin^2 2\theta^{}_{23} \left [ \alpha \left ( s^{2}_{12} \left ( s^{2}_{13} + \frac{1}{2} \sin^2 2\tilde{\theta}^{}_{13} A^{}_{CC} \right ) - c^{2}_{12} \right ) \right. \nonumber\\[3mm]
& & + \frac{1}{2} \left ( R^{}_{b} \sin2\tilde{\theta}^{}_{13} \cos2\tilde{\theta}^{}_{13} - \frac{1}{2} T^{}_{b} \sin^2 2\tilde{\theta}^{}_{13} + T^{}_{c} \right ) A^{}_{NC} \nonumber\\[3mm]
& & \left. + \frac{1}{4} T^{}_{11} \sin^2 2\tilde{\theta}^{}_{13} \left ( A^{}_{NC} - A^{}_{CC} \right ) \right ] \sin(1+A^{}_{CC})\Delta^{}_{31} \cos(C \Delta^{}_{31}) \nonumber\\[3mm]
& & - \Delta^{}_{31} \sin^2 2\theta^{}_{23} \left [ \alpha \left ( s^{2}_{12} s^{2}_{13} \frac{1+A^{}_{CC}}{C} + c^{2}_{12} \cos2\tilde{\theta}^{}_{13} \right ) \right. \nonumber\\[3mm]
& & \left. - \frac{1}{2} \left ( R^{}_{b} \sin2\tilde{\theta}^{}_{13} + T^{}_{c} \cos2\tilde{\theta}^{}_{13} \right ) A^{}_{NC} \right ] \cos(1+A^{}_{CC})\Delta^{}_{31} \sin(C \Delta^{}_{31}) \nonumber\\[3mm]
& & - 2 \Delta^{}_{31} c^{4}_{23} \sin^2 2\tilde{\theta}^{}_{13} \left [ \alpha s^{2}_{12} \frac{1 - A^{}_{CC} \cos2\theta^{}_{13}}{C} - \left ( R^{}_{b} \sin2\tilde{\theta}^{}_{13} + \frac{1}{2} T{}_{b} \cos2\tilde{\theta}^{}_{13} \right ) A^{}_{NC} \right. \nonumber\\[3mm]
& & \left. \left. + \frac{1}{2} T^{}_{11} \cos2\tilde{\theta}^{}_{13} \left ( A^{}_{NC} - A^{}_{CC} \right ) \right ] \sin(C \Delta^{}_{31}) \cos(C \Delta^{}_{31}) \right \} \; , \\[3mm]
\tilde{P}^{}_{e \mu} & \approx & \frac{1}{1 - \left ( s^{2}_{15} + s^{2}_{16} + s^{2}_{25} + s^{2}_{26} \right )} \left  \{ \left | \hat{s}^{*}_{15} \hat{s}^{}_{25} + \hat{s}^{*}_{16} \hat{s}^{}_{26} \right |^2 + \left [ s^{2}_{23} \sin^2 2\tilde{\theta}^{}_{13} \left ( 1 - T^{}_{11} - T^{}_{22} \right ) \sin(C \Delta^{}_{31}) \right. \right. \nonumber\\[3mm]
& & +  2 \alpha s^{2}_{12} s^{2}_{23} \sin^2 2\tilde{\theta}^{}_{13} \left ( \cos2\tilde{\theta}^{}_{13} \frac{A^{}_{CC}}{C} \sin(C \Delta^{}_{31}) - \Delta^{}_{31} \frac{1 - A^{}_{CC} \cos2\theta^{}_{13}}{C} \cos(C \Delta^{}_{31}) \right )  \nonumber\\[3mm]
& & + \frac{\alpha \sin2\theta^{}_{12} \sin2\theta^{}_{23} s^{}_{13}}{C A^{}_{CC}} \left ( \cos\delta \left [ \sin(1+A^{}_{CC})\Delta^{}_{31} - \frac{1 - A^{}_{CC} \cos2\theta^{}_{13}}{C} \sin(C \Delta^{}_{31}) \right ] \right. \nonumber\\[3mm]
& & \left. - \sin\delta \left [ \cos(1+A^{}_{CC})\Delta^{}_{31} - \cos(C \Delta^{}_{31}) \right ] \right ) \nonumber\\[3mm]
& & + 2 s^{}_{23} \sin2\tilde{\theta}^{}_{13} \left ( R^{}_{12} \cos2\tilde{\theta}^{}_{13} \sin(C \Delta^{}_{31}) + I^{}_{12} \cos(C \Delta^{}_{31}) \right ) \nonumber\\[3mm]
& & - \frac{A^{}_{NC} \sin2\theta^{}_{23}}{A^{}_{CC}} \left ( \sin2\tilde{\theta}^{}_{13} \left [ R^{}_{a} \left ( \sin(1+A^{}_{CC})\Delta^{}_{31} - \frac{1 - A^{}_{CC}}{C} \sin(C \Delta^{}_{31}) \right ) \right. \right. \nonumber\\[3mm]
& & \left. + I^{}_{a} \left ( \cos(1+A^{}_{CC})\Delta^{}_{31} - \cos(C \Delta^{}_{31}) \right ) \right ] \nonumber\\[3mm]
& & + \frac{2 s^{2}_{13}}{C} \left [ T^{}_{a} \left ( \sin(1+A^{}_{CC})\Delta^{}_{31} - \frac{1 + A^{}_{CC}}{C} \sin(C \Delta^{}_{31}) \right ) \right . \nonumber\\[3mm]
& & \left. \left. - I^{}_{23} \left ( \cos(1+A^{}_{CC})\Delta^{}_{31} - \cos(C \Delta^{}_{31}) \right ) \right ] \right ) \nonumber\\[3mm]
& & + s^{2}_{23} \sin2\tilde{\theta}^{}_{13} \left ( - \sin4\tilde{\theta}^{}_{13} \left [ T^{}_{b} \frac{A^{}_{NC}}{C} - T^{}_{11} \frac{A^{}_{NC} - A^{}_{CC}}{C} \right ] \left [ \sin(C \Delta^{}_{31}) - \frac{1}{2} C \Delta^{}_{31} \cos(C \Delta^{}_{31}) \right ] \right. \nonumber\\[3mm]
& & \left. \left. \left. + 4 R^{}_{b} \frac{A^{}_{NC}}{C} \left [ \cos^2 2\tilde{\theta}^{}_{13} \sin(C \Delta^{}_{31}) + \frac{1}{2} C \Delta^{}_{31} \sin^2 2\tilde{\theta}^{}_{13} \cos(C \Delta^{}_{31}) \right ] \right ) \right ] \sin(C \Delta^{}_{31}) \right \} \; , \\[3mm]
\tilde{P}^{}_{e \tau} & \approx & \frac{1}{1 - \left ( s^{2}_{15} + s^{2}_{16} + s^{2}_{35} + s^{2}_{36} \right )} \left  \{ \left | \hat{s}^{*}_{15} \hat{s}^{}_{35} + \hat{s}^{*}_{16} \hat{s}^{}_{36} \right |^2 + \left [ c^{2}_{23} \sin^2 2\tilde{\theta}^{}_{13} \left ( 1 - T^{}_{11} - T^{}_{22} \right ) \sin(C \Delta^{}_{31}) \right. \right. \nonumber\\[3mm]
& &+  2 \alpha s^{2}_{12} c^{2}_{23} \sin^2 2\tilde{\theta}^{}_{13} \left ( \cos2\tilde{\theta}^{}_{13} \frac{A^{}_{CC}}{C} \sin(C \Delta^{}_{31}) - \Delta^{}_{31} \frac{1 - A^{}_{CC} \cos2\theta^{}_{13}}{C} \cos(C \Delta^{}_{31}) \right )  \nonumber\\[3mm]
& & - \frac{\alpha \sin2\theta^{}_{12} \sin2\theta^{}_{23} s^{}_{13}}{C A^{}_{CC}} \left ( \cos\delta \left [ \sin(1+A^{}_{CC})\Delta^{}_{31} - \frac{1 - A^{}_{CC} \cos2\theta^{}_{13}}{C} \sin(C \Delta^{}_{31}) \right ] \right. \nonumber\\[3mm]
& & \left. - \sin\delta \left [ \cos(1+A^{}_{CC})\Delta^{}_{31} - \cos(C \Delta^{}_{31}) \right ] \right ) \nonumber\\[3mm]
& & + 2 c^{}_{23} \sin2\tilde{\theta}^{}_{13} \left ( R^{}_{13} \cos2\tilde{\theta}^{}_{13} \sin(C \Delta^{}_{31}) + I^{}_{13} \cos(C \Delta^{}_{31}) \right ) + R^{}_{23} \sin2\theta^{}_{23} \sin^2 2\tilde{\theta}^{}_{13} \sin(C \Delta^{}_{31}) \nonumber\\[3mm]
& & + \frac{A^{}_{NC} \sin2\theta^{}_{23}}{A^{}_{CC}} \left ( \sin2\tilde{\theta}^{}_{13} \left [ R^{}_{a} \left ( \sin(1+A^{}_{CC})\Delta^{}_{31} - \frac{1 - A^{}_{CC}}{C} \sin(C \Delta^{}_{31}) \right ) \right. \right. \nonumber\\[3mm]
& & \left. + I^{}_{a} \left ( \cos(1+A^{}_{CC})\Delta^{}_{31} - \cos(C \Delta^{}_{31}) \right ) \right ] \nonumber\\[3mm]
& & + \frac{2 s^{2}_{13}}{C} \left [ T^{}_{a} \left ( \sin(1+A^{}_{CC})\Delta^{}_{31} - \frac{1 + A^{}_{CC}}{C} \sin(C \Delta^{}_{31}) \right ) \right. \nonumber\\[3mm]
& & \left. \left. - I^{}_{23} \left ( \cos(1+A^{}_{CC})\Delta^{}_{31} - \cos(C \Delta^{}_{31}) \right ) \right ] \right ) \nonumber\\[3mm]
& & + c^{2}_{23} \sin2\tilde{\theta}^{}_{13} \left ( - \sin4\tilde{\theta}^{}_{13} \left [ T^{}_{b} \frac{A^{}_{NC}}{C} - T^{}_{11} \frac{A^{}_{NC} - A^{}_{CC}}{C} \right ] \left [ \sin(C \Delta^{}_{31}) - \frac{1}{2} C \Delta^{}_{31} \cos(C \Delta^{}_{31}) \right ] \right. \nonumber\\[3mm]
& & \left. \left. \left. + 4 R^{}_{b} \frac{A^{}_{NC}}{C} \left [ \cos^2 2\tilde{\theta}^{}_{13} \sin(C \Delta^{}_{31}) + \frac{1}{2} C \Delta^{}_{31} \sin^2 2\tilde{\theta}^{}_{13} \cos(C \Delta^{}_{31}) \right ] \right ) \right ] \sin(C \Delta^{}_{31}) \right \} \; , \\[3mm]
\tilde{P}^{}_{\mu \tau} & \approx & \frac{1}{1 - \left ( s^{2}_{25} + s^{2}_{26} + s^{2}_{35} + s^{2}_{36} \right )} \left  \{ \left | \hat{s}^{*}_{24} \hat{s}^{}_{34} c^{2}_{23} - \left ( \hat{s}^{*}_{25} \hat{s}^{}_{35} + \hat{s}^{*}_{26} \hat{s}^{}_{36} \right ) s^{2}_{23} \right |^2 \right. \nonumber\\[3mm]
& & + \frac{1}{2} \sin^2 2\theta^{}_{23} \left ( 1 - T^{}_{22} - T^{}_{33} \right ) \left [ 1 - \cos(1+A^{}_{CC})\Delta^{}_{31} \cos(C \Delta^{}_{31}) \right. \nonumber\\[3mm]
& & \left. + \cos2\tilde{\theta}^{}_{13} \sin(1+A^{}_{CC})\Delta^{}_{31} \sin(C \Delta^{}_{31}) - \frac{1}{2} \sin^2 2\tilde{\theta}^{}_{13} \sin^2 (C \Delta^{}_{31}) \right ] \nonumber\\[3mm]
& & - \frac{1}{2} \alpha s^{2}_{12} \sin^2 2\theta^{}_{23} \sin^2 2\tilde{\theta}^{}_{13} \frac{A^{}_{CC}}{C} \left [ \sin(1+A^{}_{CC})\Delta^{}_{31} + \cos2\tilde{\theta}^{}_{13} \sin(C \Delta^{}_{31}) \right ] \sin(C \Delta^{}_{31}) \nonumber\\[3mm]
& & + \frac{\alpha \sin2\theta^{}_{12} \sin2\theta^{}_{23} \cos2\theta^{}_{23} s^{}_{13} \cos\delta}{A^{}_{CC}} \left [ - \frac{1+ A^{}_{CC}}{ c^{2}_{13}} \left ( 1 - \cos(1+A^{}_{CC})\Delta^{}_{31} \cos(C \Delta^{}_{31}) \right ) \right. \nonumber\\[3mm]
& & \left. + \frac{A^{2}_{CC} + 2 s^{2}_{13} A^{}_{CC} + s^{2}_{13}}{C c^{2}_{13}} \sin(1+A^{}_{CC})\Delta^{}_{31} \sin(C \Delta^{}_{31}) + \frac{1 - A^{}_{CC} \cos2\theta^{}_{13}}{C^2} \sin^2 (C \Delta^{}_{31}) \right ] \nonumber\\[3mm]
& & - \frac{\alpha \sin2\theta^{}_{12} \sin2\theta^{}_{23} s^{}_{13} \sin\delta}{C A^{}_{CC}} \left [ \cos(1+A^{}_{CC})\Delta^{}_{31} - \cos(C \Delta^{}_{31}) \right ] \sin(C \Delta^{}_{31}) \nonumber\\[3mm]
& & - \sin2\theta^{}_{23} \sin2\tilde{\theta}^{}_{13} \left [ R^{}_{c} \left ( \sin(1+A^{}_{CC})\Delta^{}_{31} + \cos2\tilde{\theta}^{}_{13} \sin(C \Delta^{}_{31}) \right ) \right. \nonumber\\[3mm]
& & \left. - I^{}_{a} \left ( \cos(1+A^{}_{CC})\Delta^{}_{31} - \cos(C \Delta^{}_{31}) \right ) \right ] \sin(C \Delta^{}_{31}) \nonumber\\[3mm]
& & +  R^{}_{23} \sin2\theta^{}_{23} \left [ \cos2\theta^{}_{23}  \left ( 1 - \cos(1+A^{}_{CC})\Delta^{}_{31} \cos(C \Delta^{}_{31}) \right ) \right. \nonumber\\[3mm]
& & \left. + \cos2\theta^{}_{23} \cos2\tilde{\theta}^{}_{13} \sin(1+A^{}_{CC})\Delta^{}_{31} \sin(C \Delta^{}_{31}) + s^{2}_{23} \sin^2 2\tilde{\theta}^{}_{13} \sin^2 (C \Delta^{}_{31}) \right ] \nonumber\\[3mm]
& & + \sin2\theta^{}_{23} I^{}_{23} \left [ \sin(1+A^{}_{CC})\Delta^{}_{31} \cos(C \Delta^{}_{31}) + \cos2\tilde{\theta}^{}_{13} \cos(1+A^{}_{CC})\Delta^{}_{31} \sin(C \Delta^{}_{31}) \right ] \nonumber\\[3mm]
& & + \alpha \Delta^{}_{31} \sin^2 2\theta^{}_{23} \left [ \left ( s^{2}_{12} s^{2}_{13} \frac{A^{2}_{CC} + 2 s^{2}_{13} A^{}_{CC} + 1}{C^2} - c^{2}_{12} \right ) \sin(1+A^{}_{CC})\Delta^{}_{31} \cos(C \Delta^{}_{31}) \right. \nonumber\\[3mm]
& & - \left ( s^{2}_{12} s^{2}_{13} \frac{1 + A^{}_{CC}}{C} + c^{2}_{12} \cos2\tilde{\theta}^{}_{13} \right ) \cos(1+A^{}_{CC})\Delta^{}_{31} \sin(C \Delta^{}_{31}) \nonumber\\[3mm]
& & \left. + \frac{1}{2} s^{2}_{12} \sin^2 2\tilde{\theta}^{}_{13} \frac{1 - A^{}_{CC} \cos2\theta^{}_{13}}{C} \sin(C \Delta^{}_{31}) \cos(C \Delta^{}_{31}) \right ] \nonumber\\[3mm]
& & + \frac{1}{2} R^{}_{a} \frac{A^{}_{NC} \sin4\theta^{}_{23} \sin2\tilde{\theta}^{}_{13}}{A^{}_{CC} c^{2}_{13}} \left [ C \left ( 1 - \cos(1+A^{}_{CC})\Delta^{}_{31} \cos(C \Delta^{}_{31}) \right ) \right. \nonumber\\[3mm]
& & \left. - \left ( A^{}_{CC} + s^{2}_{13} \right ) \sin(1+A^{}_{CC})\Delta^{}_{31} \sin(C \Delta^{}_{31}) - c^{2}_{13} \frac{1 - A^{}_{CC}}{C} \sin^2 (C \Delta^{}_{31}) \right ] \nonumber\\[3mm]
& & + T^{}_{a} \frac{A^{}_{NC} \sin4\theta^{}_{23}}{A^{}_{CC} c^{2}_{13}} \left [ \left ( A^{}_{CC} + s^{2}_{13} \right ) \left ( 1 - \cos(1+A^{}_{CC})\Delta^{}_{31} \cos(C \Delta^{}_{31}) \right ) \right. \nonumber\\[3mm]
& & - \frac{2 s^{4}_{13} + A^{}_{CC} - 3 A^{}_{CC} \cos2\theta^{}_{13} + 2 A^{}_{CC}}{2 C} \sin(1+A^{}_{CC})\Delta^{}_{31} \sin(C \Delta^{}_{31}) \nonumber\\[3mm]
& & \left. - \frac{1}{4} \sin^2 2\tilde{\theta}^{}_{13} \left ( 1 + A^{}_{CC} \right ) \sin^2 (C \Delta^{}_{31}) \right ] \nonumber\\[3mm]
& & - \frac{A^{}_{NC} \sin2\theta^{}_{23} \sin2\tilde{\theta}^{}_{13}}{A^{}_{CC} c^{2}_{13}} \left ( I^{}_{a} c^{2}_{13} - \frac{1}{2} I^{}_{23} \sin2\theta^{}_{13} \right ) \left [ \cos(1+A^{}_{CC})\Delta^{}_{31} - \cos(C \Delta^{}_{31}) \right ] \sin(C \Delta^{}_{31}) \nonumber\\[3mm]
& & - \left ( \left [ R^{}_{b} \cos2\tilde{\theta}^{}_{13} - \frac{1}{2} T^{}_{b} \sin2\tilde{\theta}^{}_{13} \right ] \frac{A^{}_{NC}}{C} + \frac{1}{2} T^{}_{11} \sin2\tilde{\theta}^{}_{13} \frac{A^{}_{NC} - A^{}_{CC}}{C} \right ) \sin^2 2\theta^{}_{23} \sin2\tilde{\theta}^{}_{13} \nonumber\\[3mm]
& & \cdot \left [ \sin(1+A^{}_{CC})\Delta^{}_{31} + \cos2\tilde{\theta}^{}_{13} \sin(C \Delta^{}_{31}) \right ] \sin(C \Delta^{}_{31}) \nonumber\\[3mm]
& & + \frac{1}{2} \Delta^{}_{31} R^{}_{b} A^{}_{NC} \sin^2 2\theta^{}_{23} \sin2\tilde{\theta}^{}_{13} \left [ \cos2\tilde{\theta}^{}_{13} \sin(1+A^{}_{CC})\Delta^{}_{31} \cos(C \Delta^{}_{31}) \right. \nonumber\\[3mm]
& & \left. + \cos(1+A^{}_{CC})\Delta^{}_{31} \sin(C \Delta^{}_{31}) - \sin^2 2\tilde{\theta}^{}_{13} \sin(C \Delta^{}_{31}) \cos(C \Delta^{}_{31}) \right ] \nonumber\\[3mm]
& & - \frac{1}{4} \Delta^{}_{31} \left [ T^{}_{b} A^{}_{NC} - T^{}_{11} \left ( A^{}_{NC} - A^{}_{CC} \right ) \right ] \sin^2 2\theta^{}_{23} \sin^2 2\tilde{\theta}^{}_{13} \nonumber\\[3mm]
& & \cdot \left [ \sin(1+A^{}_{CC})\Delta^{}_{31} + \cos2\tilde{\theta}^{}_{13} \sin(C \Delta^{}_{31}) \right ] \cos(C \Delta^{}_{31}) \nonumber\\[3mm]
& & + \frac{1}{2} \Delta^{}_{31} T^{}_{c} A^{}_{NC} \sin^2 2\theta^{}_{23} \left [ \sin(1+A^{}_{CC})\Delta^{}_{31} \cos(C \Delta^{}_{31}) \right. \nonumber\\[3mm]
& & \left. \left. + \cos2\tilde{\theta}^{}_{13} \cos(1+A^{}_{CC})\Delta^{}_{31} \sin(C \Delta^{}_{31}) \right ] \right \} \; .
\end{eqnarray}
The above analytical approximations become invalid at the relatively large values of $L / E_{\nu}^{}$ or for the $\Delta m_{21}^{2}$-dominated oscillation. Note that, one cannot obtain the vacuum oscillation probabilities $P ( \stackrel{(-)}{\nu}^{}_{\alpha} \rightarrow \stackrel{(-)}{\nu}^{}_{\beta} )$ from $\tilde{P} ( \stackrel{(-)}{\nu}^{}_{\alpha} \rightarrow \stackrel{(-)}{\nu}^{}_{\beta} )$ in Eqs.~(B3) - (B10) by directly setting $V_{CC}^{} = V_{NC}^{} = 0$, because the expansion of $\tilde{\cal{H}}$ in Eqs.~(A4) - (A6) would be not appropriate in the limit of vanishing $V_{CC}^{}$ and $V_{NC}^{}$. In addition, one can obtain the corresponding antineutrino oscillation probabilities $\tilde{P}^{}_{\bar{\alpha} \bar{\beta}} \equiv \tilde{P}(\bar{\nu}^{}_{\alpha} \rightarrow \bar{\nu}^{}_{\beta})$ using the relation
\begin{eqnarray}
\tilde{P}^{}_{\bar{\alpha} \bar{\beta}} ( -\delta, \; \hat{s}^{*}_{ij}, \; -A^{}_{CC}, \; -A^{}_{NC} ) \; = \; \tilde{P}^{}_{\alpha \beta} ( \delta, \; \hat{s}^{}_{ij}, \; A^{}_{CC}, \; A^{}_{NC} ) \; .
\end{eqnarray}
Because of the existence of unitarity violation, the ``to all'' probabilities $\tilde{P}(\nu^{}_{\alpha} \rightarrow \nu^{}_{e, \mu, \tau}) \equiv \tilde{P}^{}_{\alpha e} + \tilde{P}^{}_{\alpha \mu} +\tilde{P}^{}_{\alpha \tau}$ for $\alpha = e, \mu, \tau$ are in generally not the unity, but displayed as
\begin{eqnarray}
\tilde{P}(\nu^{}_{e} \rightarrow \nu^{}_{e, \mu, \tau}) & \approx & 1 - 2 s^{2}_{14} + \sin2\tilde{\theta}^{}_{13} \left [ \left ( s^{2}_{14} - s^{2}_{24} s^{2}_{23} - s^{2}_{34} c^{2}_{23} \right ) \sin2\tilde{\theta}^{}_{13} + 2 R^{}_{b} \cos2\tilde{\theta}^{}_{13} \right. \nonumber\\[3mm]
& & \left. + R^{}_{23} \sin2\theta^{}_{23} \sin2\tilde{\theta}^{}_{13} \right ] \sin^2 (C \Delta^{}_{31}) + I^{}_{b} \sin2\tilde{\theta}^{}_{13} \sin(2 C \Delta^{}_{31}) \; , \\[3mm]
\tilde{P}(\nu^{}_{\mu} \rightarrow \nu^{}_{e, \mu, \tau}) & \approx & 1 - 2 s^{2}_{24} - \left ( s^{2}_{14} - s^{2}_{24} s^{2}_{23} - s^{2}_{34} c^{2}_{23} \right ) s^{2}_{23} \sin^2 2\tilde{\theta}^{}_{13} \sin^2 (C \Delta^{}_{31}) \nonumber\\[3mm]
& & + \frac{1}{2} \sin^2 2\theta^{}_{23} \left ( s^{2}_{24} - s^{2}_{34} \right ) \left [ 1 - \cos(1+A^{}_{CC})\Delta^{}_{31} \cos(C \Delta^{}_{31}) \right. \nonumber\\[3mm]
& & \left. + \cos2\tilde{\theta}^{}_{13} \sin(1+A^{}_{CC})\Delta^{}_{31} \sin(C \Delta^{}_{31}) \right ] \nonumber\\[3mm]
& & - \sin2\tilde{\theta}^{}_{13} \left [ R^{}_{a} \sin2\theta^{}_{23} \sin(1+A^{}_{CC})\Delta^{}_{31} + 2 R^{}_{b} s^{2}_{23} \cos2\tilde{\theta}^{}_{13} \sin(C \Delta^{}_{31}) \right. \nonumber\\[3mm]
& & \left. - I^{}_{a} \sin2\theta^{}_{23} \cos(1+A^{}_{CC})\Delta^{}_{31} + 2 I^{}_{b} s^{2}_{23} \cos(C \Delta^{}_{31}) \right ] \sin(C \Delta^{}_{31}) \nonumber\\[3mm]
& & + R^{}_{23} \sin2\theta^{}_{23} \left [ \cos2\theta^{}_{23}  \left ( 1 - \cos(1+A^{}_{CC})\Delta^{}_{31} \cos(C \Delta^{}_{31}) \right ) \right. \nonumber\\[3mm]
& & \left. + \cos2\theta^{}_{23} \cos2\tilde{\theta}^{}_{13} \sin(1+A^{}_{CC})\Delta^{}_{31} \sin(C \Delta^{}_{31}) + s^{2}_{23} \sin^2 2\tilde{\theta}^{}_{13} \sin^2 (C \Delta^{}_{31}) \right ] \nonumber\\[3mm]
& & + I^{}_{23} \sin2\theta^{}_{23} \left [ \sin(1+A^{}_{CC})\Delta^{}_{31} \cos(C \Delta^{}_{31}) \right. \nonumber\\[3mm]
& & \left. + \cos2\tilde{\theta}^{}_{13} \cos(1+A^{}_{CC})\Delta^{}_{31} \sin(C \Delta^{}_{31}) \right ] \; , \\[3mm]
\tilde{P}(\nu^{}_{\tau} \rightarrow \nu^{}_{e, \mu, \tau}) & \approx & 1 - 2 s^{2}_{34} - \left ( s^{2}_{14} - s^{2}_{24} s^{2}_{23} - s^{2}_{34} c^{2}_{23} \right ) c^{2}_{23} \sin^2 2\tilde{\theta}^{}_{13} \sin^2 (C \Delta^{}_{31}) \nonumber\\[3mm]
& & - \frac{1}{2} \sin^2 2\theta^{}_{23} \left ( s^{2}_{24} - s^{2}_{34} \right ) \left [ 1 - \cos(1+A^{}_{CC})\Delta^{}_{31} \cos(C \Delta^{}_{31}) \right. \nonumber\\[3mm]
& & \left. + \cos2\tilde{\theta}^{}_{13} \sin(1+A^{}_{CC})\Delta^{}_{31} \sin(C \Delta^{}_{31}) \right ] \nonumber\\[3mm]
& & + \sin2\tilde{\theta}^{}_{13} \left [ R^{}_{a} \sin2\theta^{}_{23} \sin(1+A^{}_{CC})\Delta^{}_{31} - 2 R^{}_{b} c^{2}_{23} \cos2\tilde{\theta}^{}_{13} \sin(C \Delta^{}_{31}) \right. \nonumber\\[3mm]
& & \left. - I^{}_{a} \sin2\theta^{}_{23} \cos(1+A^{}_{CC})\Delta^{}_{31} - 2 I^{}_{b} c^{2}_{23} \cos(C \Delta^{}_{31}) \right ] \sin(C \Delta^{}_{31}) \nonumber\\[3mm]
& & -  R^{}_{23} \sin2\theta^{}_{23} \left [ \cos2\theta^{}_{23}  \left ( 1 - \cos(1+A^{}_{CC})\Delta^{}_{31} \cos(C \Delta^{}_{31}) \right ) \right. \nonumber\\[3mm]
& & \left. + \cos2\theta^{}_{23} \cos2\tilde{\theta}^{}_{13} \sin(1+A^{}_{CC})\Delta^{}_{31} \sin(C \Delta^{}_{31}) - \left ( 2 + c^{2}_{23} \right ) \sin^2 2\tilde{\theta}^{}_{13} \sin^2 (C \Delta^{}_{31}) \right ] \nonumber\\[3mm]
& & - I^{}_{23} \sin2\theta^{}_{23} \left [ \sin(1+A^{}_{CC})\Delta^{}_{31} \cos(C \Delta^{}_{31}) \right. \nonumber\\[3mm]
& & \left. + \cos2\tilde{\theta}^{}_{13} \cos(1+A^{}_{CC})\Delta^{}_{31} \sin(C \Delta^{}_{31}) \right ] \; ,
\end{eqnarray}
where we have used the relation
\begin{eqnarray}
\tilde{P}^{}_{\beta \alpha} ( - \delta, \; \hat{s}^{*}_{ij} ) \; = \; \tilde{P}^{}_{\alpha \beta} ( \delta, \; \hat{s}^{}_{ij} ) \; ,
\end{eqnarray}
in matter of constant density.
One can find that these summed probabilities $\tilde{P}(\nu^{}_{\alpha} \rightarrow \nu^{}_{e, \mu, \tau})$ are all independent of the neutral-current potential $V^{}_{NC}$.

Considering the smallness of $s^{2}_{13}$, one can make further simplifications for these oscillation probabilities. To the order of $s^{2}_{13}$, we have
\begin{eqnarray}
C & \approx & \left | 1 - A^{}_{CC} \right | + \frac{2 A^{}_{CC}}{\left | 1 - A^{}_{CC} \right |} s^{2}_{13} \; = \; \gamma \left [ \left ( 1 - A^{}_{CC} \right ) + \frac{2 A^{}_{CC}}{1 - A^{}_{CC}} s^{2}_{13} \right ] \; ,
\end{eqnarray}
where
\begin{eqnarray}
\gamma & \equiv & {\rm sign} \left ( 1 - A^{}_{CC} \right ) \; = \; \frac{1 - A^{}_{CC}}{\left | 1 - A^{}_{CC} \right |} \; .
\end{eqnarray}
Then we have
\begin{eqnarray}
\sin(C \Delta^{}_{31}) & \approx & \gamma \left [ \sin(1-A^{}_{CC})\Delta^{}_{31} + 2 \Delta^{}_{31} s^{2}_{13} \frac{A^{}_{CC}}{1 - A^{}_{CC}} \cos(1-A^{}_{CC})\Delta^{}_{31} \right ] \; , \nonumber\\
\cos(C \Delta^{}_{31}) & \approx & \cos(1-A^{}_{CC})\Delta^{}_{31} - 2 \Delta^{}_{31} s^{2}_{13} \frac{A^{}_{CC}}{1 - A^{}_{CC}} \sin(1-A^{}_{CC})\Delta^{}_{31} \; .
\end{eqnarray}
If we neglect the double-suppressed terms ${\cal O} (s^{2}_{13} \alpha)$ or ${\cal O} (s^{2}_{13} s^{2}_{ij})$ or ${\cal O} (s^{4}_{13})$, the more concise approximate formulas for the neutrino oscillation probabilities in Eqs.~(11) - (16) and Eqs.~(17) - (19) can be obtained.
\end{appendix}

\end{document}